\documentclass[useAMS, usenatbib]{mn2e}
\usepackage{graphicx}
\usepackage{epsfig}

\def\vhel{\ifmmode{V_{{\rm HEL}}}\else{$V_{{\rm HEL}}$}\fi}
\def\vsys{\ifmmode{V_{\rm sys}}\else{$V_{\rm sys}$}\fi}
\def\kms{\ifmmode{~{\rm km\,s}^{-1}}\else{~km~s$^{-1}$}\fi}
\def\vlsr{\ifmmode{v_{\rm lsr}}\else{$v_{\rm lsr}$}\fi}

\def\ltsim{\ifmmode\stackrel{<}{_{\sim}}\else$\stackrel{<}{_{\sim}}$\fi}
\def\gtsim{\ifmmode\stackrel{>}{_{\sim}}\else$\stackrel{>}{_{\sim}}$\fi}

\def\reff@jnl#1{{\rm#1\/}}

\def\aj{\reff@jnl{AJ}}                  
\def\araa{\reff@jnl{ARA\&A}}            
\def\apj{\reff@jnl{ApJ}}                
\def\apjl{\reff@jnl{ApJ}}               
\def\apjs{\reff@jnl{ApJS}}              
\def\ao{\reff@jnl{Appl.Optics}}         
\def\apss{\reff@jnl{Ap\&SS}}            
\def\aap{\reff@jnl{A\&A}}               
\def\aapr{\reff@jnl{A\&A~Rev.}}         
\def\aaps{\reff@jnl{A\&AS}}             
\def\azh{\reff@jnl{AZh}}                        
\def\baas{\reff@jnl{BAAS}}              
\def\jrasc{\reff@jnl{JRASC}}            
\def\memras{\reff@jnl{MmRAS}}           
\def\mnras{\reff@jnl{MNRAS}}            
\def\pra{\reff@jnl{Phys.Rev.A}}         
\def\prb{\reff@jnl{Phys.Rev.B}}         
\def\prc{\reff@jnl{Phys.Rev.C}}         
\def\prd{\reff@jnl{Phys.Rev.D}}         
\def\prl{\reff@jnl{Phys.Rev.Lett}}      
\def\pasp{\reff@jnl{PASP}}              
\def\pasj{\reff@jnl{PASJ}}              
\def\qjras{\reff@jnl{QJRAS}}            
\def\skytel{\reff@jnl{S\&T}}            
\def\solphys{\reff@jnl{Solar~Phys.}}    
\def\sovast{\reff@jnl{Soviet~Ast.}}     
\def\ssr{\reff@jnl{Space~Sci.Rev.}}     
\def\zap{\reff@jnl{ZAp}}                        
\def\nat{\reff@jnl{Nature}}             

\def\LaTeX{L\kern-.36em\raise.3ex\hbox{a}\kern-.15em
    T\kern-.1667em\lower.7ex\hbox{E}\kern-.125emX}

\title[Source subtraction for the extended Very Small Array and 33-GHz source count estimates]{Source subtraction for the extended Very Small Array and 33-GHz source count estimates}
\author[K.A. Cleary et al.]{Kieran A. Cleary,$^1$\thanks{Present address: Jet Propulsion Laboratory, 4800 Oak Grove Drive, Pasadena CA 91109; Email: Kieran.A.Cleary@jpl.nasa.gov}, Angela C. Taylor,$^2$ Elizabeth Waldram,$^2$ Richard A. Battye,$^1$ 
\newauthor Clive Dickinson,$^1$\thanks{Present address: California Institute of Technology, 1200 E. California Blvd., Pasadena CA 91125} Rod D. Davies,$^1$ Richard J. Davis,$^1$ Ricardo Genova-Santos,$^3$ 
\newauthor Keith Grainge,$^2$ Michael E. Jones,$^2$ R\"udiger Kneissl,$^2$ G. G. Pooley,$^2$ Rafael Rebolo,$^3$ 
\newauthor Jos\'{e} Alberto Rubi\~no-Martin,$^3$ Richard D.E. Saunders,$^2$ Paul F. Scott,$^2$ 
\newauthor An\v ze Slosar,$^{2}$\thanks{Present address: Faculty of Mathematics \& Physics, University of Ljubljana, 1000 Ljubljana, Slovenia} David Titterington,$^2$ Robert A. Watson$^1$\thanks{Present address: Instituto de Astrof\'{i}sica de Canarias, 38200 La Laguna, Tenerife, Spain}\\
$^1$Jodrell Bank Observatory, University of Manchester,
  Macclesfield, Cheshire SK11 9DL \\
$^2$Astrophysics Group, Cavendish Laboratory, University of Cambridge, Madingley Road, Cambridge CB3 0HE \\
$^3$Instituto de Astrofis\'{i}ca de Canarias, 38200 La
  Laguna, Tenerife, Canary Islands, Spain}

\begin{document}

\date{}

\pagerange{\pageref{firstpage}--\pageref{lastpage}} \pubyear{2004}

\maketitle

\label{firstpage}

\begin{abstract}
We describe the source subtraction strategy and observations for the extended Very Small Array, a CMB interferometer operating at 33~GHz. A total of 453 sources were monitored at 33~GHz using a dedicated source subtraction baseline. 131 sources brighter than 20~mJy were directly subtracted from the VSA visibility data. Some characteristics of the subtracted sources, such as spectra and variability, are discussed. The 33-GHz source counts are estimated from a sample selected at 15~GHz. The selection of VSA fields in order to avoid bright sources introduces a bias into the observed counts. This bias is corrected and the resulting source count is estimated to be complete in the flux-density range 20--114~mJy. The 33-GHz source counts are used to calculate a correction to the VSA power spectrum for sources below the subtraction limit.

\end{abstract}


\section{Introduction}
Extragalactic radio sources are a major contaminant of Cosmic Microwave Background (CMB) observations, particularly at small angular scales. The power spectrum (${\cal C}_{\ell} \equiv \ell(\ell+1)C_{\ell}/2\pi$) of Poissonian-distributed point sources increases as $\ell^{2}$ whereas that of the CMB decreases with increasing $\ell$ due to Silk damping and the incoherent addition of fluctuations along the line of sight through the surface of last scattering. The point-source foreground will therefore dominate CMB measurements at high $\ell$ in the absence of an effective source-subtraction strategy.

Three factors complicate the implementation of schemes for dealing with point sources. Firstly, in the frequency range $\approx$ 30--200~GHz where many CMB experiments operate, no large-area surveys for sources exist. Simple extrapolation of flux densities from low-frequency data can cause worse source contamination than if no subtraction were attempted at all, due to the population of sources with rising spectra \citep{taylor_etal_01}. The positions of sources from low-frequency catalogues may be used to `project-out' \citep*{bond_etal_98} sources from the high-frequency ($\gtsim 30$~GHz) CMB data, but at a cost of signal-to-noise. Secondly, `flat'-spectrum (flux-density spectral index $\alpha \approx 0$, $S \propto \nu^{-\alpha}$) sources will increasingly dominate bright flux density-selected samples at higher frequencies. Flat-spectrum sources are compact objects and often show flux density variations which means that accurate source subtraction at higher frequencies requires monitoring of source flux densities simultaneously with CMB observations. Thirdly, in order to avoid confusion by the CMB, surveying for sources should be performed at high resolution where the CMB contribution is negligible. 

The Very Small Array (VSA) is a collaboration between the Cavendish Astrophysics Group (University of Cambridge), the University of Manchester's Jodrell Bank Observatory (JBO) and the Instituto de Astrof\'{\i}sica de Canarias (IAC) in Tenerife. The instrument is a 14-element interferometer operating in Ka band (26--36~GHz) and located at Teide Observatory, Tenerife \citep[see e.g.][]{paper1}. The CMB power spectrum up to $\ell \simeq 800$ has been measured by the VSA in its compact configuration \citep[][]{paper2,paper3} and up to $\ell \simeq 1500$ \citep[][]{paper5,dickinson_etal_04} in an extended configuration. A unique source subtraction strategy is employed by the VSA to allow accurate subtraction of sources. The regions observed by the VSA are surveyed with the Ryle Telescope \citep[RT; see e.g.][]{jones_91} in Cambridge at $15$~GHz \citep[see e.g.][]{waldram_etal_03}. The sources detected at 15~GHz are then observed with a single-baseline interferometer at the same frequency as the VSA {\em and\/} at the same time.

The purpose of this paper is to describe the source subtraction observations and discuss some characteristics of the subtracted sources, such as spectra and variability. The 33-GHz source counts are estimated and used to determine a correction to the VSA power spectrum for unsubtracted sources.

\section{VSA Source Subtraction Strategy}

VSA extended array measurements currently probe angular multipoles up to $\ell \simeq 1500$, where measurements of the power spectrum can be significantly affected by Poisson noise from unsubtracted radio sources. In order to mitigate this potentially serious foreground, the VSA strategy is three-fold. 

Firstly, VSA fields are selected such that there are no sources predicted to be brighter than 500~mJy at 30~GHz on the basis of the NRAO VLA $1.4$-GHz All Sky Survey \citep[NVSS;][]{condon_98} and the Green Bank $4.8$-GHz \citep[GB6;][]{gregory_96} survey. These predictions were made by extrapolating the flux density of every source in the $4.8$-GHz catalogue using the spectral index between 1.4~GHz and 4.8~GHz. 

Secondly, the RT in Cambridge is used to survey the VSA fields at 15~GHz in advance of CMB observations. The RT synthesised beam FWHM for this purpose is 25 arcsec corresponding to a spherical multipole, $\ell \approx 5 \times 10^{4}$, where the contribution from the CMB is negligible. The area surveyed for a typical VSA extended array field is some $13$~deg$^{2}$, whereas the area within the FWHM of the RT primary beam is only 0.01~deg$^{2}$. This means that a raster scanning technique must be used so that the $15$-GHz survey can keep up with the VSA observations. \cite{waldram_etal_03} describe the $15$-GHz observations and present source lists.

Thirdly, the sources detected at 15~GHz are monitored at 33~GHz simultaneously with CMB observations using a dedicated source subtraction baseline.  

The depth of the RT survey is determined by the 33-GHz source subtraction flux density limit, $S_{\rm lim}$, which in turn is prescribed by the requirement that the confusion noise from unsubtracted sources is at best very small compared with the thermal noise on a VSA map. The confusion noise, $\sigma_{\rm conf}$, can be estimated as \citep{Scheuer57}
\begin{equation}
\centering
\sigma^{2}_{\rm conf} = \Omega \int_{0}^{S_{\rm lim}} S^{2}\,n(S)\,dS,
\label{conf_noise_1.eqn}
\end{equation}
\noindent where the differential source count, $n(S)$\,d$S$, is defined as the number of sources per steradian in the flux density range $S$ to $S$+d$S$, $\Omega$ is the VSA synthesised beam solid angle and sources with flux densities greater than $S_{\rm lim}$ have been subtracted. Then, from equation~\ref{conf_noise_1.eqn}, for a given confusion noise, the required $S_{\rm lim}$ is given by
\begin{equation}
\centering
S_{\rm lim} = \left [\frac{\sigma_{\rm conf}^{2}(3-\beta)}{\kappa\:S_{0}^{\beta}\:\Omega} \right ]^{1/(3-\beta)},
\label{conf_noise.eqn}
\end{equation}
\noindent where the differential source count is parameterised as $n(S) = \kappa\;(S/S_{0})^{- \beta}$. 

At the outset, the differential source counts were not known at the VSA observing frequency and the source counts from RT observations for the Cosmic Anisotropy Telescope ~\citep[CAT;][]{robson_etal_93} were used. \cite{osullivan_etal_95} found an integral count N($S>10$~mJy) $\approx$ 3 per deg$^{2}$ at 15~GHz and and $\beta$ was taken to be approximately 1.8. The thermal noise on a compact array VSA map is $\approx 30$~mJy beam$^{-1}$ and the compact VSA synthesised beamwidth is $\approx 30$ arcmin. Assuming a conservative average spectral index between 15 and 34~GHz, $\bar{\alpha}_{15}^{34} = 0$, the source subtraction limit such that the VSA compact array maps are thermal noise dominated is $S_{\rm lim} \ltsim 109$~mJy. In fact, a subtraction limit of 80~mJy was used for the compact array. In order that sources with rising spectral indices as extreme as $\alpha_{15}^{34} =-2$ are found at 5 $\sigma$ at $15$~GHz, the sensitivity of the RT survey was $\sigma \approx$ 4~mJy.

For the extended array observations, which measure the power spectrum at higher multipoles, a deeper RT survey is required. Using the updated counts from the compact array source monitoring, we have $n(S)=54\;S^{-2.15}$ Jy$^{-1}$ sr$^{-1}$~\citep{paper2}. The nominal r.m.s.\ noise on a VSA extended array map is $\approx 5$~mJy beam$^{-1}$, the synthesised beamwidth is $\approx 11$ arcmin and the required $S_{\rm lim}$ is $\approx 20$~mJy at 33~GHz. This means that the sensitivity of the RT survey for the extended array must be at least $\approx$ 0.8~mJy in order to find inverted-spectrum sources with $\alpha_{15}^{33} =-2$ at 5 $\sigma$. In fact, the requirement that the RT must survey the VSA fields in advance of VSA observations means that a practical limit to the RT sensitivity is reached at 2~mJy. As a consequence, not all sources with rising spectral indices $\alpha_{15}^{33} \ltsim -1$ will be found and there remains a possibility that some such sources exist in the VSA fields. Clearly, for measurements at higher $\ell$, a new strategy is required

\section{Observations}

\begin{figure}
\begin{center}
\includegraphics[width=0.5\textwidth,angle=0]{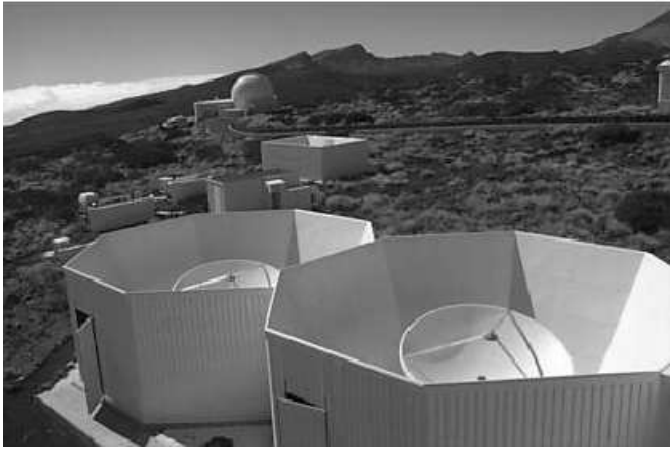}
\caption[VSA source subtractor dishes adjacent to the VSA at Teide Observatory, Tenerife.]{VSA source subtractor dishes in individual enclosures to minimise ground spill-over. The dishes are aligned North-South, have diameters of 3.7 m and are separated by 9.2 m.}
\label{ssub.fig}
\end{center}
\end{figure}

The VSA source subtraction system consists of two 3.7-m dishes spaced 9.2 m apart, used as a single-baseline North-South interferometer located adjacent to the main array (Fig.~\ref{ssub.fig}). The receivers, IF system and correlator are identical to those for the VSA. Each dish is located within an enclosure to minimize ground spill-over. With this baseline, the interferometer is not sensitive to large-scale CMB emission, while the contribution on smaller angular scales is expected to be negligible. Both the VSA and the source subtraction system measure the same combination of Stokes parameters. Observations of each source were made over a range of hour angles similar to those used by the main array.

Over the course of the extended array monitoring programme, the source subtractor dishes were replaced with ones of higher surface accuracy and the system was upgraded to double sideband operation. This increased the aperture efficiency from $\approx 0.5$ to $\approx 0.7$ and doubled the effective bandwidth. The nominal r.m.s.\ of visibility amplitude for the source subtractor following the upgrade is $\sigma \approx 0.34$~Jy~s$^{1/2}$.

\begin{table}
\centering
 \caption{Nominal specifications of the VSA source subtractor.}
  \begin{tabular}{lc}
    \hline
Location       			&Izana, Tenerife (2340 m) \\
Maximum declination range 	&$-5^{\circ}< $Dec $< +60^{\circ}$ \\
No. of antennas 		&2 \\
Baseline separation		&9.2~m\\
Frequency range			&$26-36$ GHz \\
System temperature, $T_{\rm sys}$ (K)	&$\approx 35$ K \\
Bandwidth, $\Delta \nu$		&1.5 GHz \\
Sidebands			&two\\
Correlator			&complex\\
Dish diameter			&3.7~m\\
Point source flux sensitivity (1-$\sigma$) & $\approx 0.34$ Jy s$^{1/2}$ \\ \hline
\label{specs} 
 \end{tabular}
\end{table}

\begin{figure}
\begin{center}
\includegraphics[width=0.48\textwidth,angle=0]{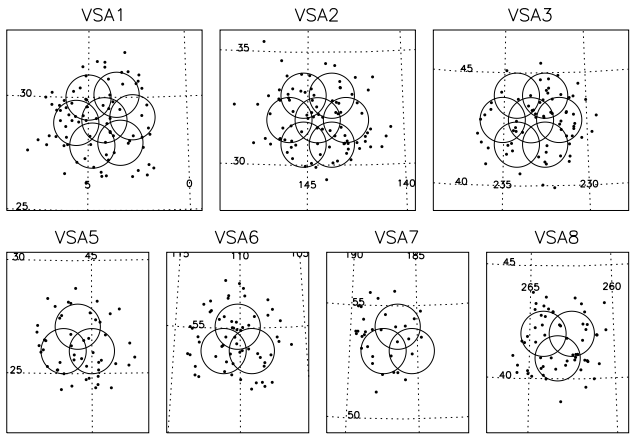}
\caption[]{The points indicate the locations of the 453 sources monitored for each of the VSA mosaics. The $2\degr$ FWHM of each VSA pointing is indicated by a circle. The projection is gnomonic and coordinates of RA(J2000) and Dec(J2000) are in degrees.}
\label{fields_new.fig}
\end{center}
\end{figure}

The purpose of the monitoring programme is to measure the flux densities of the sources detected by the RT at 15~GHz at the VSA observing frequency of 33~GHz. This takes place simultaneously with CMB observations by the VSA. For a typical observation, the source subtractor drive system takes the coordinates of the sources within a radius of some $2^{\circ}$ of the current field centre from the $15$-GHz source list. A subset (typically $\approx 30$) of these sources is monitored over the course of each $\approx 5$-hour observing run, with interleaved observations of the brightest source in the field for phase calibration. In this manner, the flux densities of the sources are sampled over the course of the observations of a given field. The integration time for each sample is 400 s, giving a nominal sensitivity per sample of 17~mJy and typically $\approx$ 30 samples are obtained for each source. A total of 453 sources were monitored for the purposes of source subtraction in 7 independent regions as shown in Fig.~\ref{fields_new.fig}.

\section{Data Reduction and Calibration}

Data reduction for the source subtractor is performed in the same manner as for the main array data, using the {\sc reduce} software package \citep[see e.g.][]{paper1}. The calibrated visibility time-series for each source in a given VSA field is examined by eye to check the quality of the phase and flux-density calibration. Bad data are flagged at this point and the samples for each source are averaged over the observing period of that field by the main array. 

Due to the much greater instantaneous sensitivity of the source subtractor compared to the VSA, the brightest calibrators used for the VSA (Tau\,A, Cas\,A and Jupiter) are not suitable for the source subtractor. In addition, Tau\,A and Cas\,A are resolved by the source subtraction baseline. The main flux density calibrator for the source subtractor is NGC~7027, a planetary nebula with a flux density of $5.45 \pm  0.2$ Jy at 32~GHz, based on a measurement by \cite{mason_etal_99}. Assuming a spectral index, $\alpha = 0.1 \pm 0.1$, a flux density of 5.43 Jy at 33~GHz is adopted. The calibration uncertainty in the source flux densities is 4 per cent due to the uncertainty in the \cite{mason_etal_99} measurement. 

As with the main array, a baseline-based amplitude calibration is applied. Since the source subtractor slews between pointed observations, interleaved observations of phase calibrator sources are made. The phase calibrator sources need only be bright enough to achieve good signal-to-noise ratios in a 400-s integration and are typically the brightest sources already identified in each VSA field.

\section[]{Source Subtraction for Extended VSA Data}
\label{ssub_evsa}

As described in \cite{dickinson_etal_04}, a campaign of high-sensitivity measurements of the CMB  up to $\ell \simeq 1500$ was conducted with the VSA in an extended configuration. The flux densities of the sources detected at 15~GHz in the VSA fields were monitored by the source subtractor at 33~GHz during observations by the main array, as described above. 

\begin{figure*}
\begin{center}
\includegraphics[width=1.0\textwidth, angle=0]{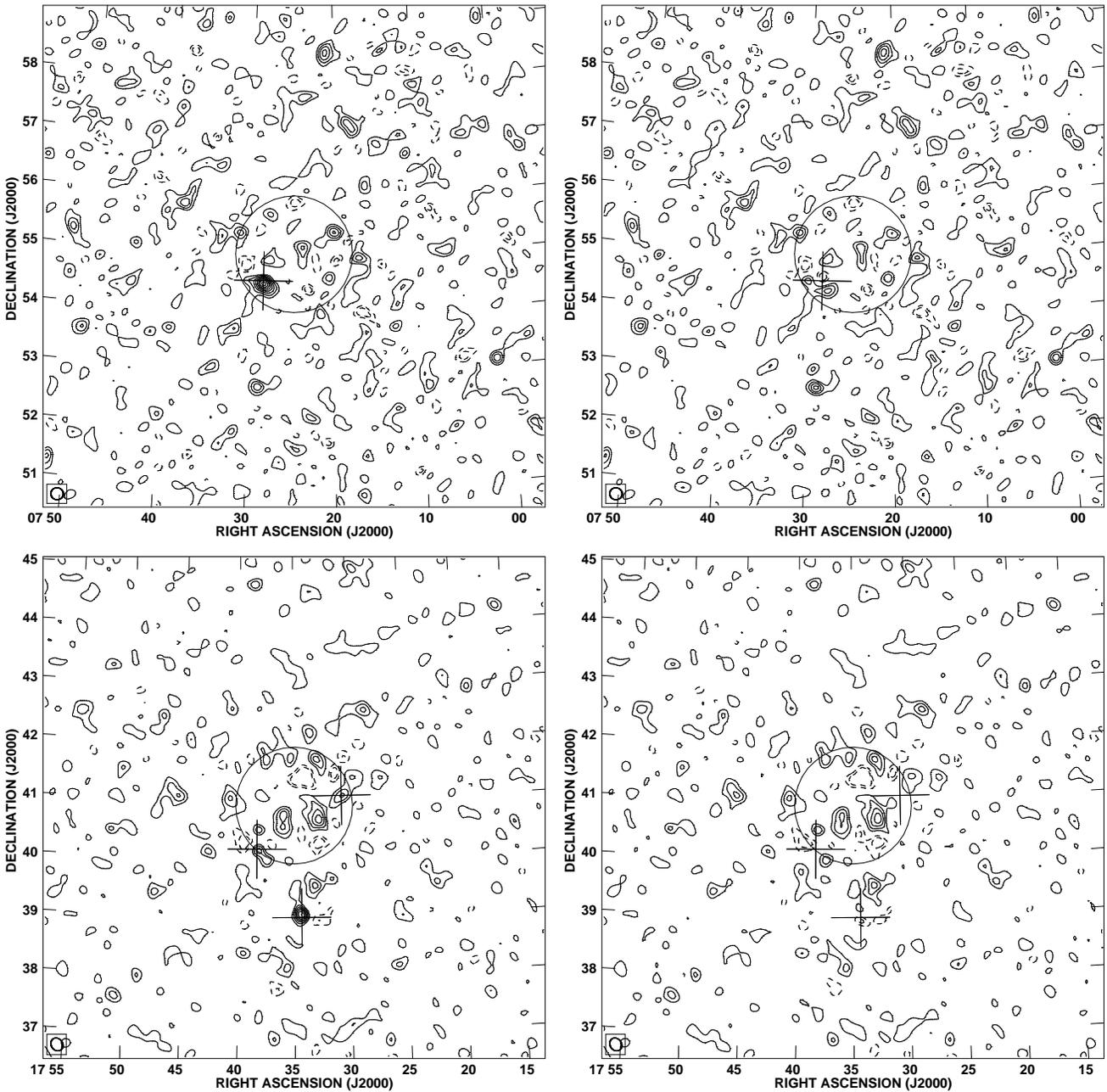}
\caption[]{Maps of the VSA6G (top row) and VSA8G (bottom row) fields before (left) and after (right) source subtraction. The crosses mark the positions of the sources predicted to have the brightest apparent flux density from source subtractor data (Table~\ref{app_flux.table}). The circles in both maps indicate the area under the primary beam $2\degr$ FWHM. Contour levels are -30, -20, 10, 20, 30, 40, 50, 60, 70, 80, 90, 100 and 110~mJy/beam.}
\label{ss_maps.fig}
\end{center}
\end{figure*}

For each field, the sources found to have mean flux densities over the observation period greater than 20~mJy were subtracted directly from the VSA visibility data. Fig.~\ref{ss_maps.fig} shows two examples of a VSA field before and after source subtraction, demonstrating successful source subtraction and revealing CMB structure previously confused by the presence of bright sources.

A total of 131 sources were found to have flux densities greater than 20~mJy and were directly subtracted from the visibility data. Of these 131 sources, the brightest apparent flux density on a VSA map was 115~mJy (see Table~\ref{app_flux.table}).

Some characteristics of the monitored sources, such as spectra and variability, are now discussed.

\begin{table}
\centering
\caption[]{The sources subtracted from the VSA6G and VSA8G fields in Fig.~\ref{ss_maps.fig}. The flux density measured by the source subtractor, $S$, and the apparent (i.e.\ attenuated by the primary beam, as on a VSA map) flux density, $S_{\rm app}$, are given.}
\begin{tabular}{lccccc}
\hline
Field&RA(J2000)&Dec(J2000)&$S$ [Jy]&$S_{\rm app}$ [Jy]\\
\hline
VSA6G&07$^{\rm h}$ 28$^{\rm m}$ 23$\fs$8&+54$\degr$ 31$\arcmin$ 16$\farcs$8&0.154&0.115\\
VSA8G&17$^{\rm h}$ 34$^{\rm m}$ 20$\fs$6&+38$\degr$ 57$\arcmin$ 51$\farcs$4&0.995&0.089\\
     &17$^{\rm h}$ 30$^{\rm m}$ 41$\fs$4&+41$\degr$ 02$\arcmin$ 56$\farcs$6&0.069&0.045\\
     &17$^{\rm h}$ 38$^{\rm m}$ 19$\fs$1&+40$\degr$ 08$\arcmin$ 19$\farcs$6&0.101&0.056\\
\hline
\end{tabular}
\label{app_flux.table}
\end{table}

\subsection{Source spectra}
\label{ssub_spectra}

\begin{figure*}
\begin{center}
\includegraphics[width=1.0\textwidth, angle=0]{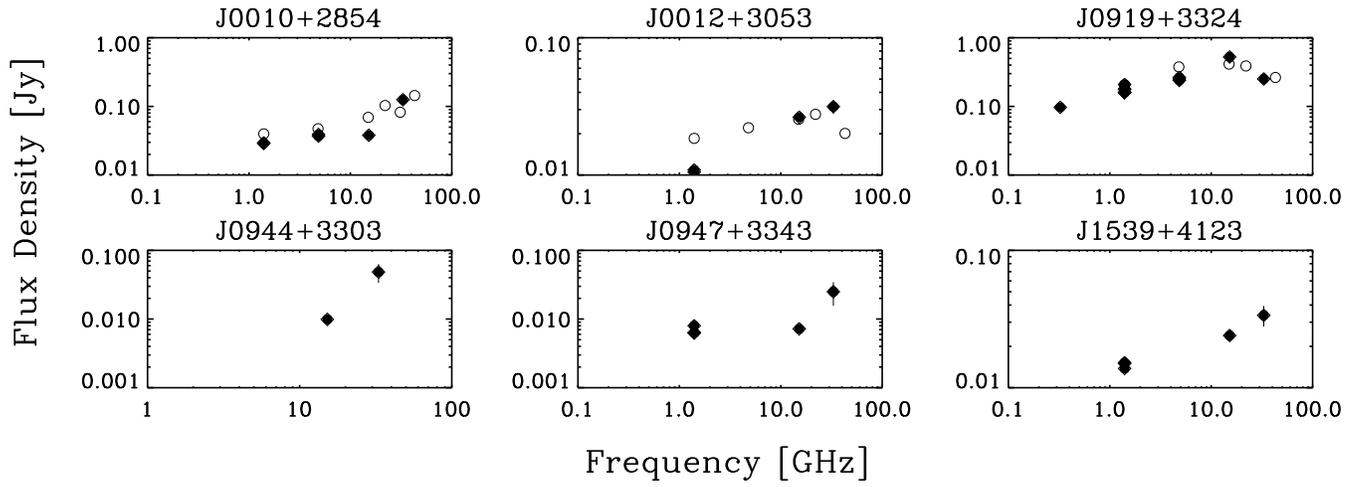}
\caption[]{Examples of sources with rising spectra, on the basis of the 15 and 33-GHz flux densities as well as the flux densities of positional counterparts in the CATS database (filled circles). Data from measurements by B04 are also plotted, where available (unfilled circles). The source J0944+3303 is the only source monitored at 33~GHz for source subtraction not to have a positional counterpart in NVSS or FIRST and has a spectral index of -2 between 15 and 33~GHz.}
\label{inv.fig}
\end{center}
\end{figure*}

In order to examine the spectra of subtracted sources, a catalogue search was performed for the brightest counterparts within 10 arcsec using the Astrophysical Catalogues Support System \citep[CATS;][]{Verkhodanov_etal_97}. Fig.~\ref{inv.fig} shows some examples of the inverted-spectrum sources including J0944+3303, the only subtracted source not to have a positional counterpart in NVSS or FIRST, with a spectral index $\alpha = -2$ between 15 and 33~GHz. Where available, data from measurements made over a range of frequencies simultaneously \citep[][hereafter B04]{bolton_etal_04} are also plotted.

In order to assess whether any of the subtracted sources were significantly peaked, an attempt was made to fit the spectra to a broken power-law with the following formula \citep{marecki_etal_99}:

\begin{equation}
\centering
S(\nu)=\frac{S_{0}}{1-e^{-1}} (\nu/\nu_{0})^{k}(1-e^{-(\nu/\nu_{0})^{l-k}}),
\label{marecki.eqn}
\end{equation}

{\noindent where $k$ and $l$ are the spectral indices of the rising and falling parts of the spectrum respectively and $S_{0}$ and $\nu_{0}$ are fitting parameters. In some cases, the spectrum was significantly peaked but a second-order polynomial of the form,}

\begin{equation}
\centering
\log S_{\nu} = S_{0} + \alpha \log \nu - c (\log \nu)^{2}
\label{marecki2.eqn}
\end{equation}

\begin{figure*}
\begin{center}
\includegraphics[width=0.9\textwidth, angle=0]{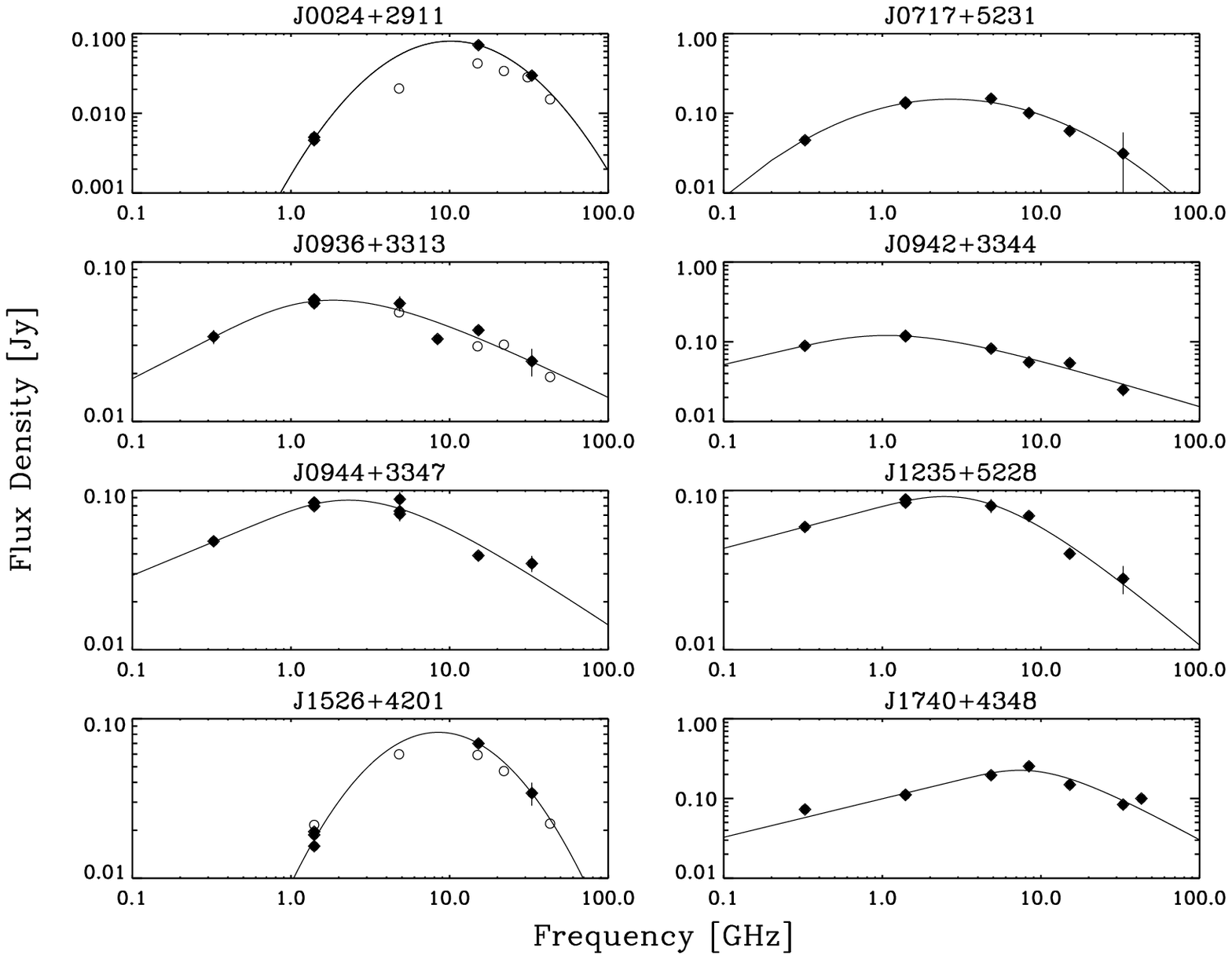}
\caption[]{Spectra for peaked-spectrum candidate sources on the basis of the 15 and 33-GHz flux densities as well as the flux densities of positional counterparts in the CATS database (filled circles). Data from measurements by B04 are also plotted, where available (unfilled circles). The solid curves are broken power-law or second-order polynomial fits to the data.}
\label{gps.fig}
\end{center}
\end{figure*}

{\noindent provided a better fit to the data. Fig.~\ref{gps.fig} shows the 8 sources where equation~\ref{marecki.eqn} provided a good fit to the available data, or the curvature, $c$, of the second-order polynomial fit was $> 0.5$. Table~\ref{spec_fit.table} gives the associated fitting parameters.}

There is considerable interest in the identification of gigahertz peaked (GPS) sources and those sources with spectra peaking at frequencies above a few gigahertz, known as high frequency peakers (HFP). The 9C survey \citep{waldram_etal_03} is the first radio-frequency survey to cover an appreciable area at a frequency above the 4.8-GHz GB6 survey with a sensitivity down to $\approx 10$~mJy. As such, it allows HFP candidates to be selected on the basis of their spectra between 4.8 and 15~GHz. The follow-up observations of a subset of these sources at 33~GHz presented here provide another high-frequency datum. An attempt has been made here to investigate if any of the subtracted sources are significantly peaked on the basis of 15-GHz and 33-GHz data as well as low-frequency data from catalogues. However, source variability means that follow-up observations at a range of frequencies simultaneously, such as those performed by B04, may be required to verify the existence of spectral peaks. Of the 8 sources determined here to be peaked-spectrum, 3 sources were also observed by B04. Of these 3 sources, J0024+2911 and J1526+4201 were also found to be peaked by B04 and their measurements of the third source, J0936+3313, are well fit by the falling portion of the broken power-law, although they lack lower-frequency flux-density measurements to verify the presence of a peak around 1~GHz. This suggests that a high-frequency survey combined with a catalogue search is a useful method of selecting peaked-spectrum candidates, despite the variability of sources.

\begin{table}
\centering
\caption[]{Spectral parameters for peaked sources of Fig.~\ref{gps.fig}. The parameters are, peak flux density, $S_{\rm max}$, and corresponding frequency, $\nu_{\rm max}$, the spectral indices of the rising ($k$) and falling ($l$) parts of the broken power-law fit, and the curvature, $c$, of the second-order polynomial fit.}
\label{spec_fit.table}
\begin{tabular}{lccccc}
\hline
Name(J2000)&$S_{\rm max}$ [Jy]&$\nu_{\rm max}$ [GHz]&$k$&$l$&$c$\\
\hline
0024+2911&0.081&10.2&-   &-    &1.6\\
0717+5231&0.151&2.7 &-   &-    &0.6\\
0936+3313&0.058&1.8 &-0.51&0.48&-\\
0942+3344&0.120&1.1 &-0.47&0.59&-\\
0944+3347&0.087&2.3 &-0.41&0.67&-\\
1235+5228&0.091&2.5 &-0.26&0.83&-\\
1526+4201&0.083&8.3 &-   &-    &1.1\\
1740+4348&0.226&7.4 &-0.48&1.09&-\\
\hline
\end{tabular}
\end{table}

\subsection[]{Source variability}

For the majority of sources, the source subtractor lacks sufficient sensitivity for measurements of day-to-day flux-density variability, since we use it to measure the average flux density over the time-scale of the VSA observations. However, we can get an idea of the range of variability in the sources monitored for source subtraction in two ways. Firstly, the VSA fields overlap so that the sources lying in the overlap are observed at different epochs. By comparing the mean flux density measurements at different epochs, the variability of sources in the over lap regions can be examined. Secondly, a $\chi^{2}$ test was applied to find those sources whose time-series could not be fitted by a constant flux density. This latter method is only useful for bright sources observed frequently (e.g. phase calibrators) which thus have sufficient signal-to-noise.

\begin{figure}
\begin{center}
\includegraphics[scale=0.45]{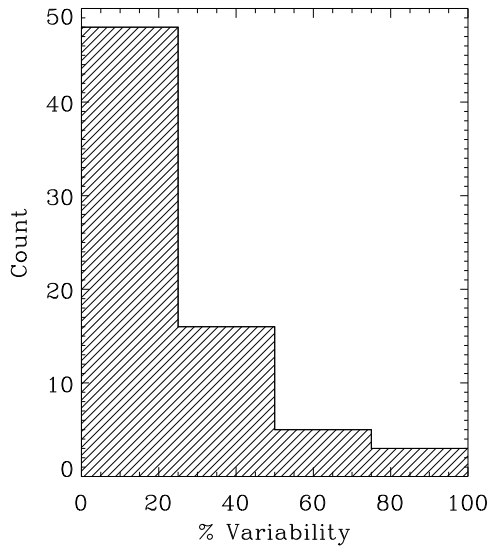}
\caption{Histogram showing the number of sources in the overlapping areas of the VSA fields whose maximum variability, expressed as a percentage of mean flux density, falls in the ranges 0--25, 26--50, 51--75 and 76--100 per cent.}
\label{var_hist.fig}
\end{center}
\end{figure}

In order to compare the flux densities measured at different epochs, the flux densities measured at 34.1~GHz were interpolated to $33$~GHz. Fig.~\ref{var_hist.fig} shows a histogram of the percentage variability of the 72 sources in the overlapping regions with flux densities greater than 20 mJy. Of these 72 sources, 23 were found to have varied by more than 25 per cent of the mean flux density, 8 by more than 50 per cent and 3 by more than 75 per cent.

The above analysis does not take into account the error in the flux density measurements at different epochs. In order to determine whether a source is variable or not, a $\chi^{2}$ test is applied, i.e.

\begin{equation}
\centering
\chi_{r}^{2} = \frac{1}{N-1}\sum_{i=1}^{N}\left( \frac{S_{i}-\langle S \rangle}{\Delta S_{i}}\right )^{2},
\label{chi_sq.eqn}
\end{equation}

\begin{figure*}
\begin{center}
\includegraphics[width=0.9\textwidth, angle=0]{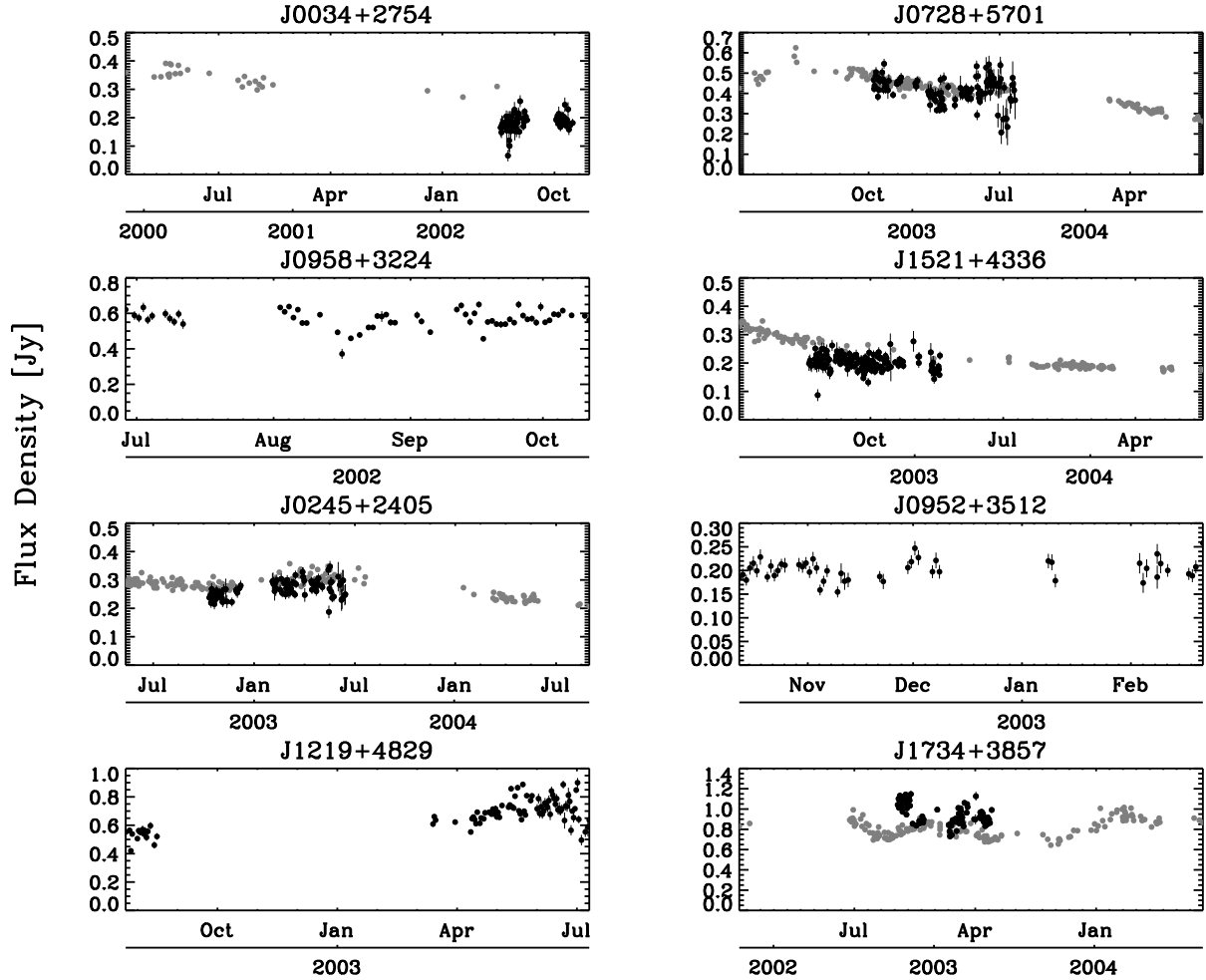}
\caption{The measured 33~GHz (black filled circles) and 15~GHz (grey filled circles) flux density time-series for variable sources (as determined by the $\chi^{2}$ test). 15~GHz data were not available for all sources. All 8 sources were used as phase calibrators by the source subtractor.}
\label{var_sources_only.fig}
\end{center}
\end{figure*}

{\noindent where N is the number of measurements, the $S_{i}$ are the individual flux densities and $\Delta S_{i}$ their errors. This tests the hypothesis that the flux density time-series could be modelled by a constant. Sources for which this probability is $\le 0.1 \%$ are taken to be variable. Of the 453 sources monitored at 30~GHz for source subtraction purposes, only 8 were found to be variable according to this criteriona and these were all phase calibrators. Fig.~\ref{var_sources_only.fig} shows the measured flux density time-series (as well as the 15~GHz time-series, where available) for the 8 sources found to be variable using the $\chi^{2}$ test. While these sources are taken to be variable for the purposes of further analysis, that is not to suggest that other sources in the sample are definitely not variable.}

\begin{figure*}
\begin{center}
\includegraphics[width=0.9\textwidth, angle=0]{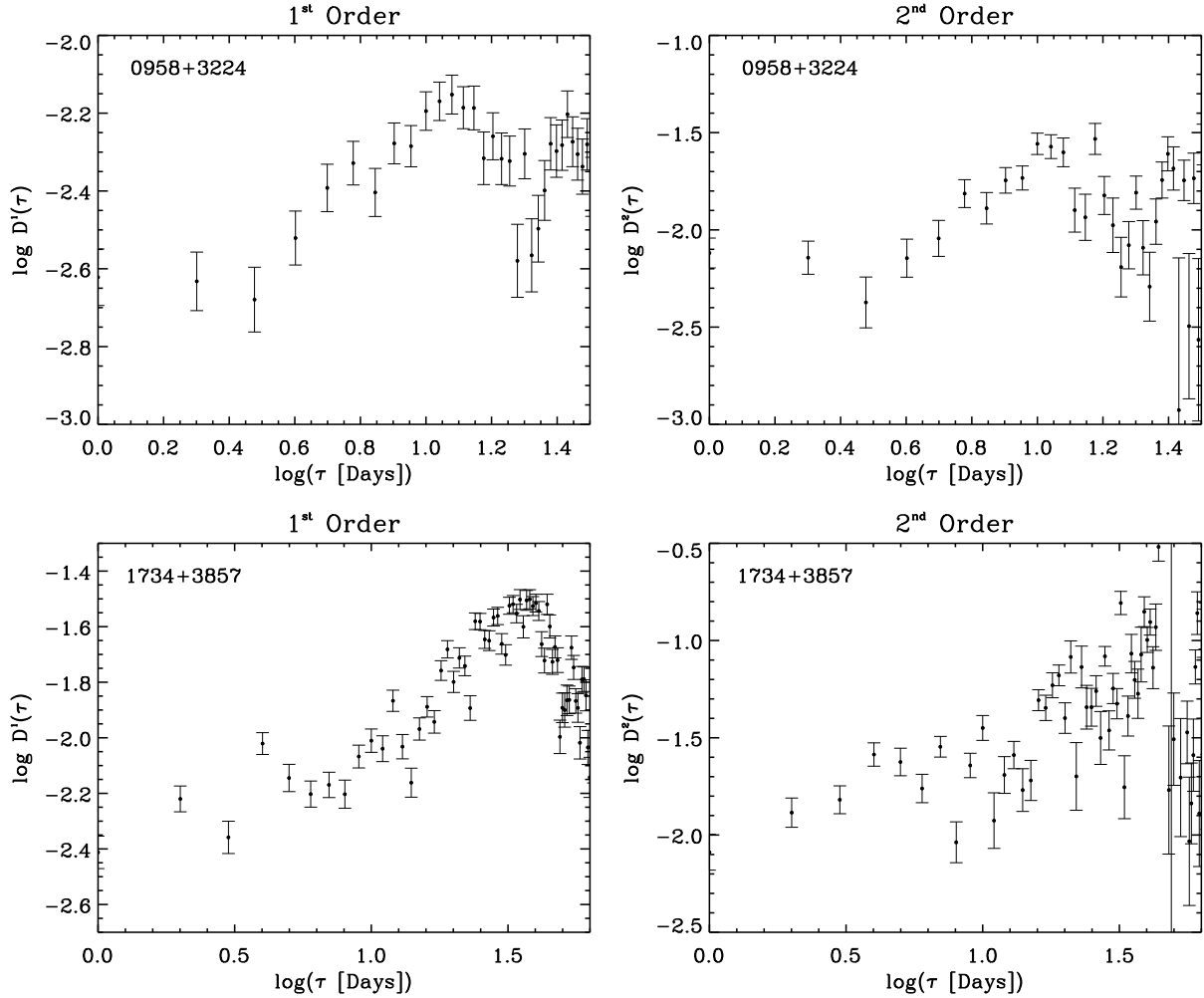}
\caption{The first and second-order structure functions for J0958+3224 (top left and right) and J1734+3857 (bottom left and right).}
\label{struct_fns.fig}
\end{center}
\end{figure*}

For these 8 sources, the structure functions \citep[see e.g. ][]{simonetti_etal_85} were estimated in order to check for the presence of characteristic timescales in the flux density variations. Structure function analysis allows time-variability to be quantified without the problems encountered using Fourier techniques with irregularly sampled data. The first-order and second-order structure functions are defined as
\begin{eqnarray}
\centering
D^{1}(\tau)&=&\langle [S(t+\tau)-S(t)]^{2} \rangle\nonumber
\label{struct_fns.eqn}
\end{eqnarray}
\noindent and
\begin{eqnarray}
\centering
D^{2}(\tau)&=&\langle [S(t+2\tau)-2S(t+\tau)+S(t)]^{2} \rangle \nonumber 
\label{struct_fns1.eqn}
\end{eqnarray}
\noindent respectively, where, $S(t)$ is the source flux density at time, $t$, $\tau$ is the `lag' and the angle brackets denote an ensemble average. For a stationary random process, the structure functions are related to the variance of the process, $\sigma^{2}$, and its autocorrelation function, $\rho(\tau)$, as follows:
\begin{eqnarray}
\centering
D^{1}(\tau)&=&2\sigma^{2} [1+\rho(\tau)],\nonumber  \\
D^{2}(\tau)&=&6\sigma^{2} [1+\frac{1}{3}\rho(2\tau)+\frac{4}{3}\rho(\tau)]. \nonumber 
\label{struct_fns2.eqn}
\end{eqnarray}
For $\tau \ll \tau_{0}$, where $\tau_{0}$ is the smallest correlation time-scale present in the series, the structure function, $D^{M}(\tau)$ has a logarithmic slope $2M$. For $\tau \gg \tau_{1}$, where $\tau_{1}$ is the greatest correlation time-scale present in the series, the structure functions reach a plateau of $2\sigma^{2}$ for $M=1$ and $6\sigma^{2}$ for $M=2$. For intermediate time-scales, the logarithmic slope of the structure function is $\leq 2M$. Well-defined structure functions were found for 2 of the 8 sources taken to be variable and these are plotted in Fig.~\ref{struct_fns.fig}. We now comment on each of these sources.

\noindent {\bf J0958+3224} (3C232) -- A compact steep-spectrum quasar with $z = 0.53$. The first-order structure function for this source has a well-defined slope of $\approx 0.8$ on time-scales of 3--12 days, turning over at $\tau \approx 12$ days. The second-order structure function has a slope of $\approx 1.3$ in the same time-scale range, indicating that day-like structure exists in the time-series since purely linear trends (or segments of longer-term variations) would produce a flat second-order structure function. The time-series (Fig.~\ref{var_sources_only.fig}) shows variations of up to $\approx 30$ per cent.

\noindent {\bf J1734+3857} -- This source is identified with the flat-spectrum BL\,Lac object NVSS J173420+385751 at $z = 0.97$. The well-defined logarithmic slope of $\approx 0.9$ in the first-order structure function on time-scales of 6--39 days indicates the presence of structure in the time-series with a maximum correlation time-scale of $\approx 39$ days. The second-order structure-function is less clear, however the slope of $\approx 0.9$ over the same time-scales appears inconsistent with the presence of purely linear trends. The time-series (Fig.~\ref{var_sources_only.fig}) shows variations of up to $\approx 20$ per cent.

\begin{figure}
\begin{center}
\includegraphics[scale=0.58, angle=0]{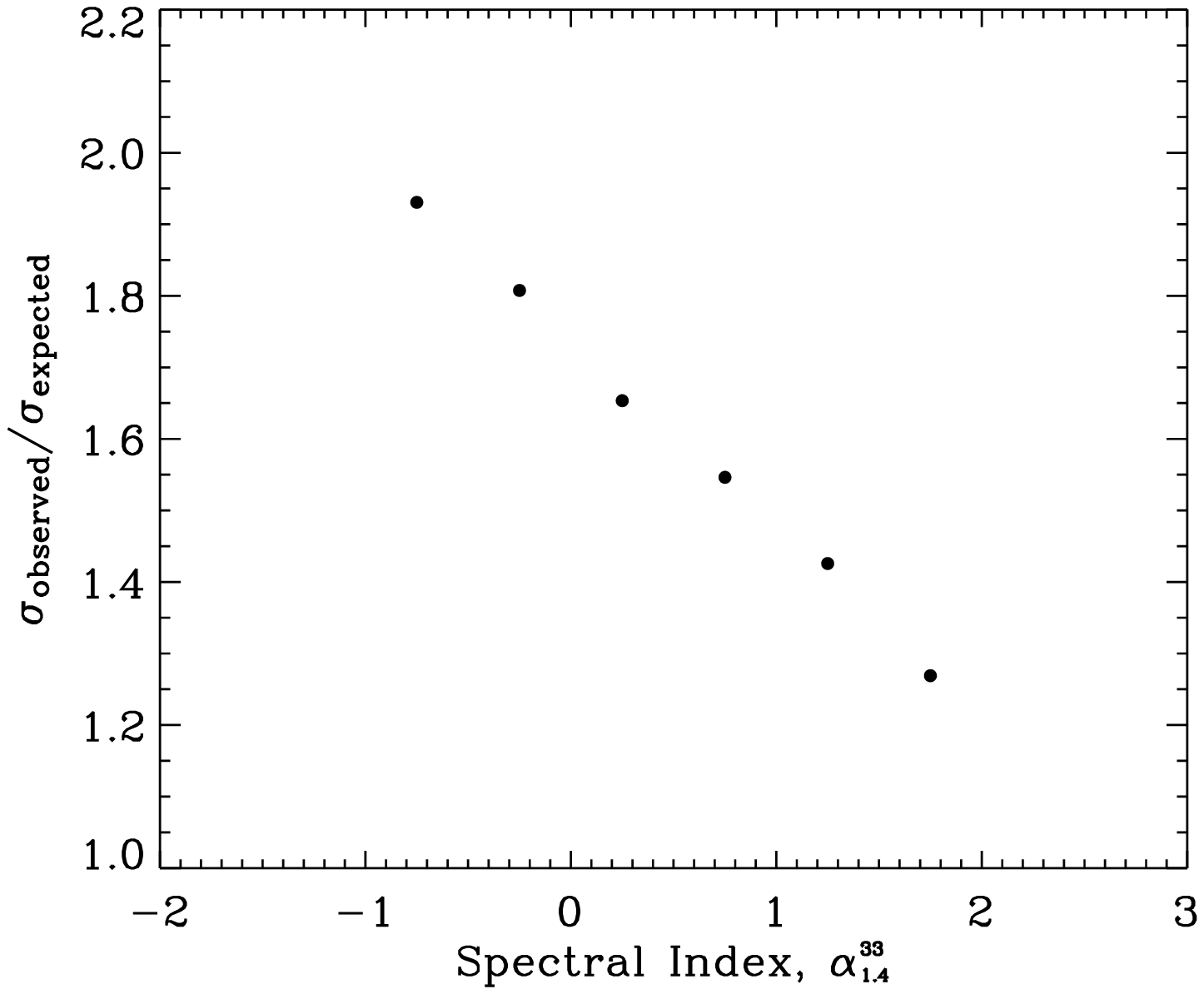}
\caption[]{The ratio of observed-to-expected r.m.s.\ of flux density, for all 453 sources monitored at 33~GHz for the purposes of source subtraction, averaged in bins of $\Delta \alpha_{1.4}^{33}=0.5$. The ratio of the observed-to-expected r.m.s.\ rises with decreasing spectral index suggesting that the excess scatter in the flux density time-series is due to variability associated with flat and rising-spectrum sources.}
\label{dex_var.fig}
\end{center}
\end{figure}

Although the $\chi^{2}$ test indicates variability in 8 sources, these are amongst the 11 brightest sources monitored for source subtraction and so have been observed frequently as phase calibrators. Fainter sources may be variable but will not be identified so easily as such by a $\chi^{2}$ test due to the decreased signal-to-noise ratio of these observations. Fig.~\ref{dex_var.fig} plots the ratio of observed-to-expected r.m.s.\ of flux density, for all 453 sources observed for source subtraction against the spectral index between 1.4 and 33~GHz, averaged in bins of $\Delta \alpha_{1.4}^{33}=0.5$. If differences between the observed and expected r.m.s.\ of flux density are due to variability, then we might expect the ratio of observed-to-expected r.m.s.\ to increase with decreasing spectral index on the basis that the majority of variable compact sources are flat-spectrum, and this appears to be the case.

\section[]{33-GHz Source Count Estimates}
\label{ssub_src_counts}

The set of sources subtracted from VSA data does not form a well-selected sample for the purposes of source-count estimation. In order to estimate the 33-GHz source counts, a new source sample was constructed as follows. We selected well-defined regions within the existing VSA fields which had previously been surveyed at 15~GHz down to a completeness limit of 10~mJy. The selected regions encompass a total area of 0.044 steradian and 370 sources with 15-GHz flux densities greater than 10~mJy were identified.  All the sources identifed above the completeness limit at 15~GHz were observed at 33~GHz with the source subtraction baseline.  Hence the source count we present at 33~GHz is based on a sample selected at 15~GHz. In addition, since the VSA fields were specifically chosen to have no sources brighter than 500~mJy at 33~GHz (on the basis of extrapolation from NVSS and GB6 data) we do not expect our source counts to be complete at high flux densities.

\begin{figure}
\begin{center}
\includegraphics[scale=0.58, angle=0]{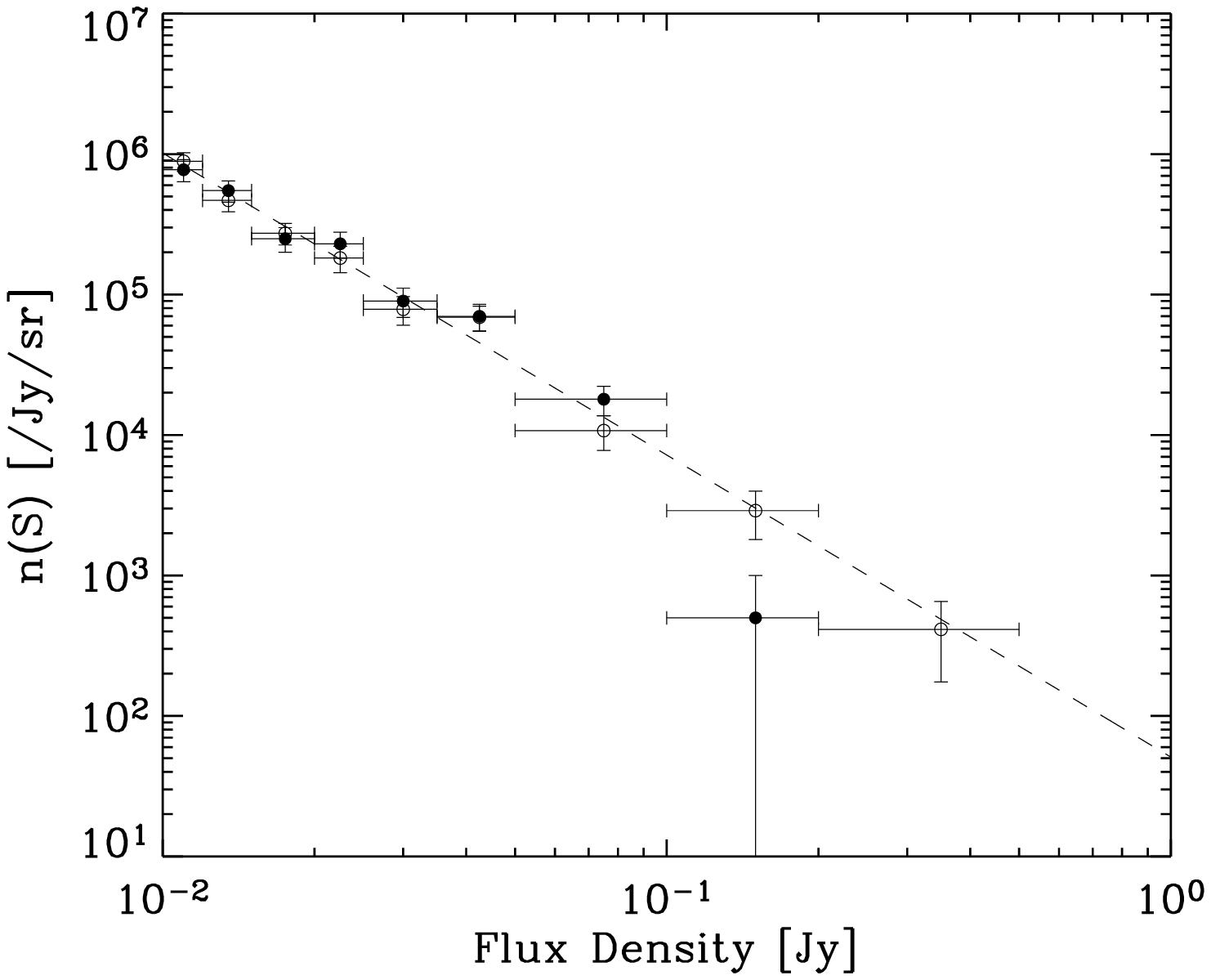}
\caption[]{15~GHz differential source counts derived from sources in VSA123 (filled circles) and VSA5678 (unfilled circles) fields. The dashed line is the known 15~GHz source count ($n(S) \approx 51\;S^{-2.15}$) from the 9C survey. There is a (deliberate) deficiency of bright sources in the VSA123 fields.}
\label{scount33_final_w_15ghz.fig}
\end{center}
\end{figure}

We have also identified a further bias in our data. The first three VSA fields (VSA1, VSA2 and VSA3, hereafter collectively VSA123) were chosen with an extra constraint, such that no source predicted to be brighter than $\approx$ 250~mJy at 33~GHz should lie within the FWHM of the primary beam. The later VSA fields (VSA5, VSA6, VSA7 and VSA8, hereafter collectively VSA5678) were chosen with the more relaxed point source limit of 500~mJy since their positions were already constrained by operational requirements. This slight difference in field selection has resulted in a (deliberate) deficiency of sources with flux densities greater than $\approx$ 100~mJy at 15~GHz in VSA123. This can be seen by comparing the differential counts at 15~GHz for the VSA123 and VSA5678 fields with that determined using the 9C survey (see Fig.~\ref{scount33_final_w_15ghz.fig}). In order to use sources from the VSA123 fields in our 33~GHz source-count estimate, we therefore attempt to take account of this bias.

\begin{figure}
\begin{center}
\includegraphics[scale=0.57, angle=0]{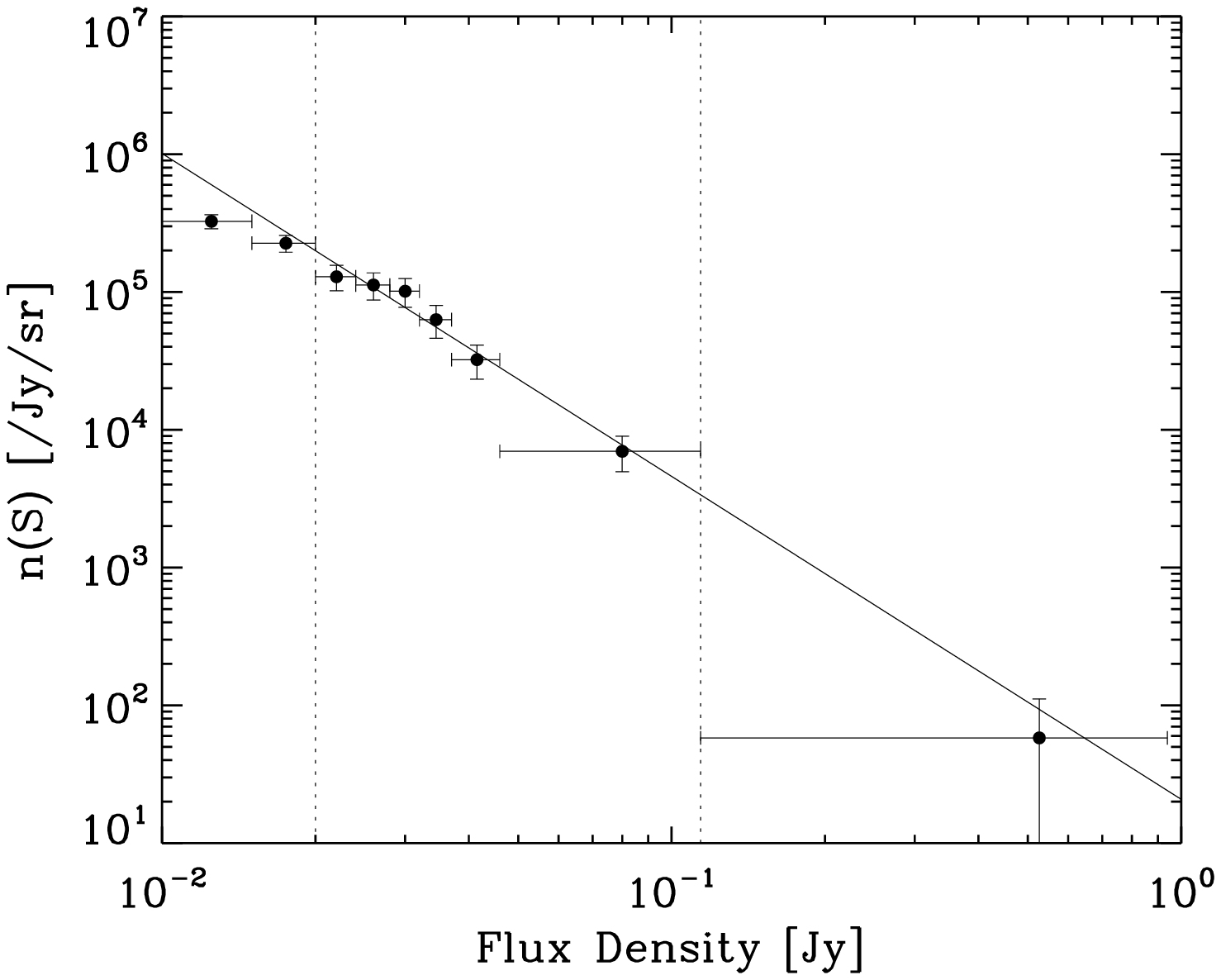}
\caption[]{The $33$-GHz differential source counts showing a fall-off in flux density bins $< 20$~mJy and $> 114$~mJy, ascribed to lack of completeness. The best-fitting power law to flux densities in the range 20--114~mJy (102 sources) is also plotted (solid line). The vertical dotted lines indicate the flux density range used in the fit. The vertical error bars indicate the Poisson errors in each bin and the horizontal error bars indicate the bin widths.}
\label{scount33.fig}
\end{center}
\end{figure}

In using sources from VSA123 to estimate the counts we need to set an upper limit, $S_{\rm upper}$, to the flux density at 33~GHz to ensure that the original selection at 15~GHz (which is incomplete above $S_{\rm 15~GHz} \approx$ 100~mJy) does not bias the count. One approach is to examine the 15--33~GHz spectral indices of all sources above 100~mJy at 15~GHz in our whole sample. We find 12 sources, with a highest (falling) spectral index of 1.0. Since this a small sample, we have investigated the spectral index distributions for complete samples at lower flux densities at 15~GHz. We find that approximately 80 per cent of these sources have $\alpha_{15}^{33} \leq 1.0$. This value of 1.0 gives us an estimate for $S_{\rm upper}$ of 46~mJy. A second approach has been to examine the 33-GHz source counts separately for VSA123 and VSA5678; we find the deficit in the VSA123 count relative to the VSA5678 count appears to become significant at approximately 80~mJy. We therefore assume that 46~mJy is a conservative estimate for $S_{\rm upper}$ and combine the counts from VSA123 and VSA5678, using only those sources from VSA123 with flux densities $\leq 46$~mJy. Fig.~\ref{scount33.fig} shows the resulting differential source counts. The data used for the fit are shown in Table~\ref{source_count.table}. A correction is made for the bin widths such that the number of sources in each bin of width, $W$, and centre, $S_{c}$, is multiplied by a factor of $(1-r^2)$ where $r=W/2S_{c}$.

Parameterising the differential source counts as a power law, 

\begin{table}
\centering
\caption[]{Table showing the data used to fit the differential source count in Fig.~\ref{scount33.fig}. The normalising area for each bin is also provided.}
\label{source_count.table}
\begin{tabular}{cccc}
\hline
Bin start&Bin width&Number&Normalising\\
$[$Jy$]$&$[$Jy$]$&in bin&area [sr]\\
\hline
0.020&0.004&23&0.044\\
0.024&0.004&20&0.044\\
0.028&0.004&18&0.044\\
0.032&0.005&14&0.044\\
0.037&0.009&13&0.044\\
0.046&0.068&14&0.024\\
\hline
\end{tabular}
\end{table}

\begin{equation}
\centering
n(S)=\kappa \left (\frac{S}{S_{0}}\right )^{-\beta}, 
\label{power_law.eqn}
\end{equation}

{\noindent the least-squares fit to the power law was determined for flux densities in the range 20--114~mJy. Choosing $S_{0} = 70$~mJy to lie in the centre of this range, the resulting fit gives}

\begin{equation}
\centering
n(S) = (10.6^{+2.3}_{-2.2}) \left(\frac{S}{70 \mbox{ mJy}}\right)^{-2.34^{+0.25}_{-0.26}} \quad\mbox{mJy$^{-1}$ sr$^{-1}$.}\quad
\label{ssub_scount.eqn}
\end{equation}

{\noindent The quoted errors correspond to the $\chi^{2}$ 1-$\sigma$ confidence limit for each parameter. Expressed in Jy$^{-1}$ sr$^{-1}$, the source count is}

\begin{equation}
\centering
n(S) \approx 21 \; S^{-2.34} \quad\mbox{Jy$^{-1}$ sr$^{-1}$,}\quad
\label{ssub_scount1.eqn}
\end{equation}

{\noindent which is also plotted in Fig.~\ref{scount33.fig}. The errors on the data points are the formal Poisson errors, $\sqrt{N}$ in $N$.}

\begin{figure}
\begin{center}
\includegraphics[scale=0.57, angle=0]{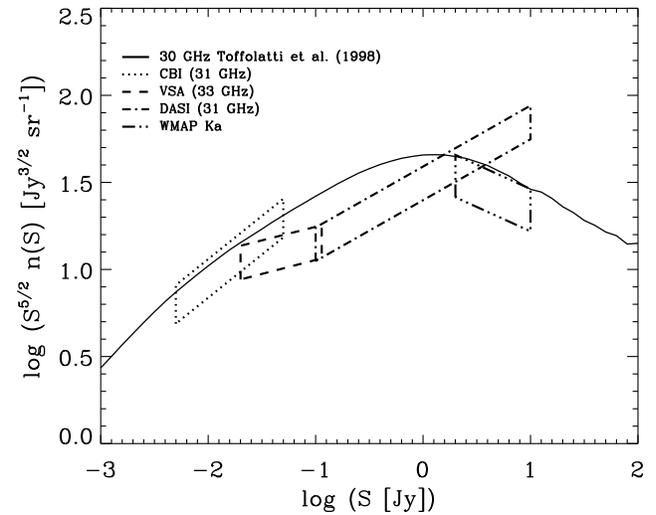}
\caption[VSA $33$-GHz differenctial source counts (normalised to those expected for a Euclidean universe) compared with those derived by CBI, DASI and WMAP.]{VSA $33$-GHz differential source counts (normalised to those expected for a Euclidean universe) compared with those derived by CBI \citep{mason_etal_03}, DASI \citep{kovac_etal_02} and WMAP \citep{bennett_etal_03}. The delineated regions for each experiment are defined by the quoted 1-$\sigma$ error on the normalisation of $n(S)$. The solid curve shows the predicted counts from the 30-GHz model of \cite{toff_98}.}
\label{diff2.fig}
\end{center}
\end{figure}

Fig.~\ref{diff2.fig} shows the differential source count predictions of a 30-GHz model \citep[][hereafter T98]{toff_98}, the source counts derived from WMAP \citep{bennett_etal_03}, CBI \citep{mason_etal_03} and DASI \citep{kovac_etal_02} experiments as well as the VSA count described by equation~\ref{ssub_scount1.eqn}. All counts are normalised to those expected in a Euclidean universe. The observed counts are consistent in the overlapping flux-density ranges, but generally fall below the T98 model. The observed VSA counts will be used to re-scale the T98 model in the following section.

\section[]{Residual Correction to the CMB Power Spectrum}

The contribution to the temperature fluctuations from faint sources below the subtraction limit, $S_{\rm lim}$, has a component due to the Poissonian-distributed sources as well as a clustering term, as follows:
\begin{eqnarray}
\centering
T_{\rm CMB}^{2} C_{\rm src}&=& \left(\frac{\partial B_{\nu}}{\partial T}\right)^{-2} \left( C_{\rm src}^{\rm Poisson} + C_{\rm src}^{\rm Clustering} \right)
\label{ssub_resid0.eqn}
\end{eqnarray}
\noindent where
\begin{equation}
\centering
\frac{\partial B_{\nu}}{\partial T} = \frac{2k}{c^2} \left(\frac{k T_{\rm CMB}}{h} \right)^{2} \frac{x^{4}e^{x}}{(e^x-1)^2},
\label{ssub_resid1.eqn}
\end{equation}
\noindent $B_{\nu}$ is the Planck function, $k$ is Boltzmann's constant and $x \equiv h\nu/kT_{\rm CMB}$. We can estimate each term of equation~\ref{ssub_resid0.eqn} using (see equation~\ref{conf_noise_1.eqn})
\begin{equation}
C_{\rm src}^{\rm Poisson}= \int_{0}^{S_{\rm lim}} S^{2}\,n(S)\,dS \\
\label{ssub_resid.eqn}
\end{equation}
\noindent and \citep{ScottWhite99}
\begin{equation}
C_{\rm src}^{\rm Clustering}= \omega_{\ell}\left( \int_{0}^{S_{\rm lim}} S\,n(S)\,dS \right)^{2},
\label{ssub_resid2.eqn}
\end{equation}
\noindent where $\omega_{\ell}$ is the Legendre transform of the angular two-point function of sources, $\omega(\theta)$. Based on estimates of the angular correlation function of sources from the NVSS and FIRST surveys \citep[see e.g.][]{blake_wall_02b}, we estimate that the contribution of the clustering term (equation~\ref{ssub_resid2.eqn}) is negligible over all scales measured by the VSA.

\begin{figure}
\begin{center}
\includegraphics[scale=0.57, angle=0]{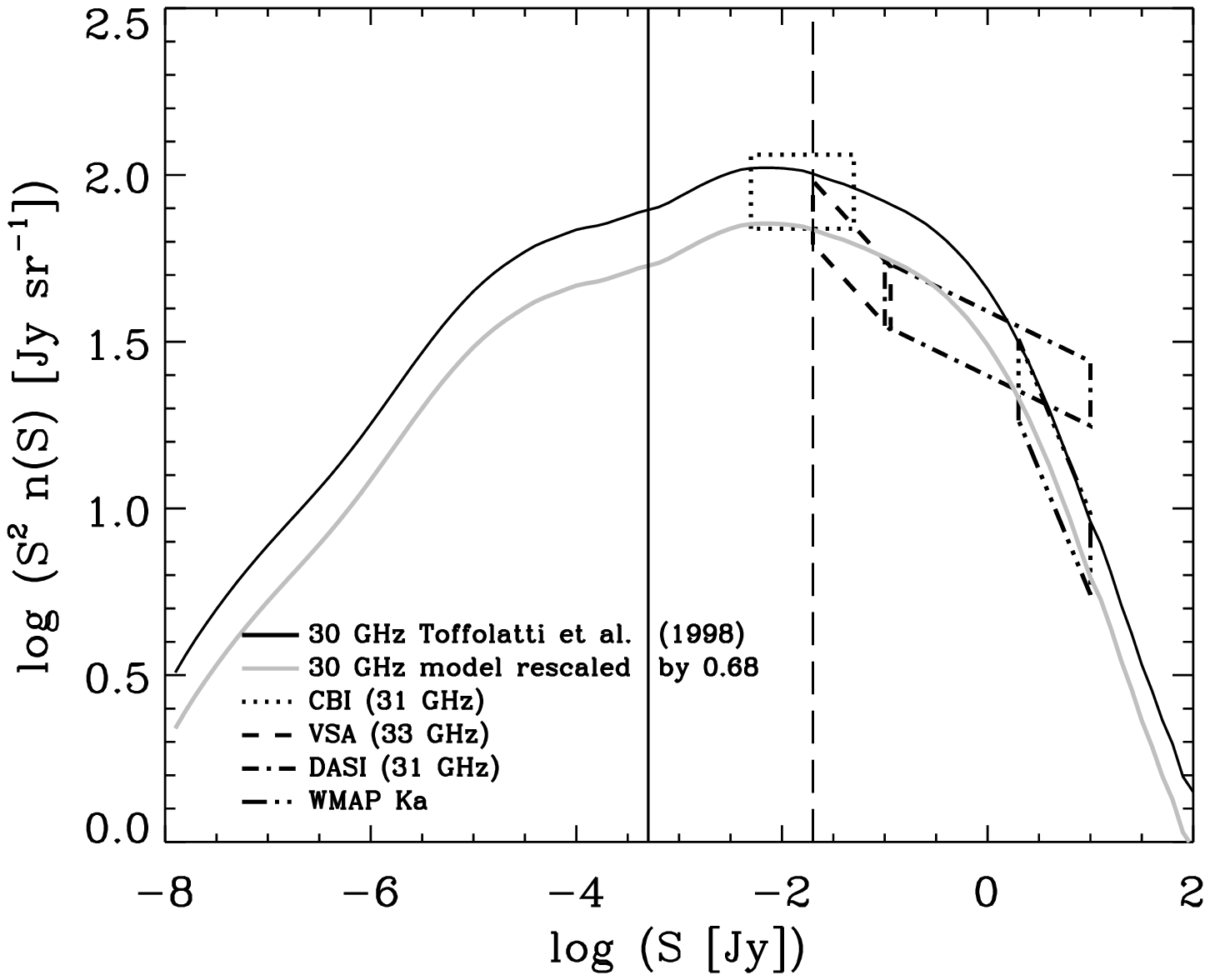}
\caption[]{$S^{2} n(S)$ for the VSA compared with those derived by CBI \citep{mason_etal_03}, DASI \citep{kovac_etal_02} and WMAP \citep{bennett_etal_03}. The delineated regions for each experiment are defined by the quoted 1-$\sigma$ error on the normalisation of n(S). The solid black curve shows the predicted counts from the model of \cite{toff_98}, while the solid grey curve is this model re-scaled by a factor of 0.68 in order to best fit the VSA data. The source subtraction limit of 20~mJy is indicated by the vertical long-dashed line. The vertical solid line indicates the lower integration limit for which the residual source correction is 99 per cent of the total.}
\label{diff4.fig}
\end{center}
\end{figure}

Fig.~\ref{diff4.fig} plots the integrand of equation~\ref{ssub_resid.eqn} on a log scale for the VSA and other experiments operating at a similar frequency. The vertical long-dashed line indicates the upper limit of integration, $S_{\rm lim} =0.02$ Jy and the vertical solid line indicates the lower limit of integration for which the residual correction is 99 per cent of the total. For differential source counts with power-law slopes $\beta < 3$, the dominant contribution to the Poisson noise comes from the flux decades just below the source subtraction limit \citep{toff_98}.

The measured source counts are used to estimate the contribution from sources below the flux density subtraction limit. However, it is clear that extrapolating the best-fit source count to lower flux densities will over-estimate the counts of faint sources since, as the T98 model shows, the counts are expected to flatten at lower flux densities. The VSA source counts were used to re-scale the T98 model and the best fit is plotted as the grey curve in Fig.~\ref{diff4.fig}. The re-scaling of the T98 model required to best fit the VSA 33~GHz source counts was found to be $0.68 \pm 0.07$. Table~\ref{resids.table} shows the residual source contribution for the VSA estimated using source counts measured by experiments operating at similar frequencies, the T98 model and that model re-scaled by 0.68 (denoted T98$^{\star}$).

\begin{table*}
\centering
\caption[]{Residual source power for a subtraction limit of 20~mJy calculated using the source counts ($n(S)=\kappa \, S^{-\beta}$ Jy$^{-1}$ sr$^{-1}$) measured by the VSA and other experiments as well as the T98 30-GHz model and that model re-scaled by 0.68 (T98$^{\star}$). The expected residual source contribution at $\ell=1000$ is also given. Due to the expected flattening of the counts at lower flux densities, source counts measured at high flux densities (e.g.\ WMAP) will over-estimate the residual source power for subtraction to low flux densities.}
\label{resids.table}
\begin{tabular}{lcccc}
\hline
 &$\kappa$&$\beta$&$T_{\rm CMB}^{2}C_{\rm src}$ [$10^{-5}$~$\umu$K$^{2}$ sr]&$(\Delta T_{\rm src})^2|_{\ell=1000}$ [$\umu$K$^{2}$]\\
\hline
VSA&20.8&2.34&229&359\\
CBI&92.0&2.00&174&278\\
DASI&32.0&2.15&128&204\\
WMAP&44.0&2.80&9532&15186\\
T98& & &191&305\\
T98$^{\star}$& & &130&207\\
\hline
\end{tabular}
\end{table*}

\begin{figure*}
\begin{center}
\includegraphics[scale=0.73, angle=90]{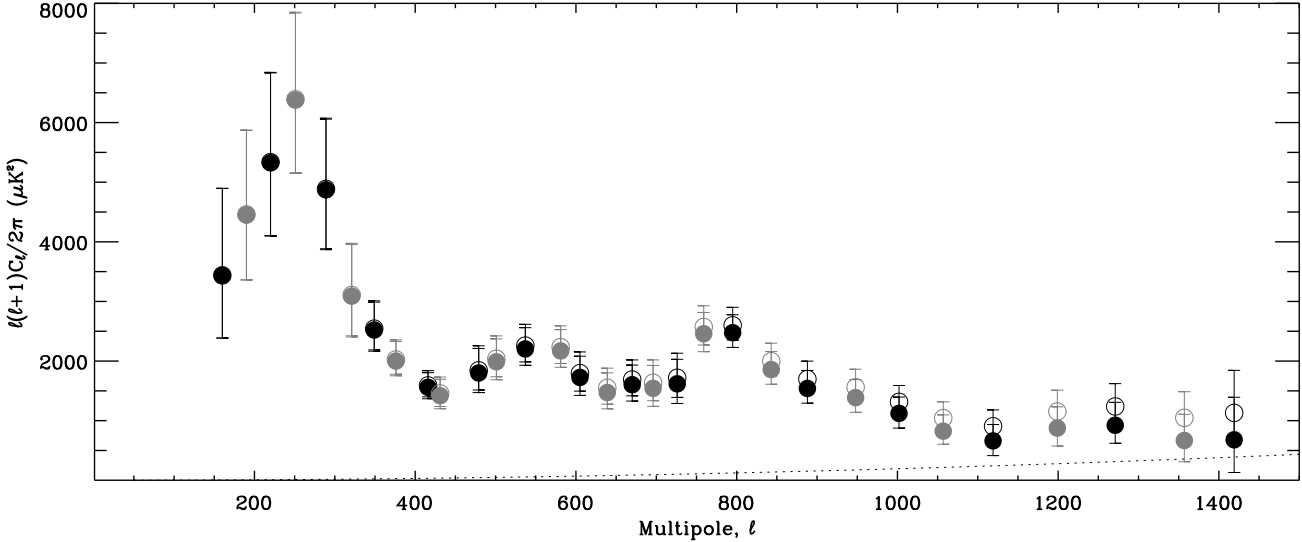}
\caption[]{The final, double-binned, VSA extended array power spectrum before (unfilled circles) and after (filled circles) the estimated residual source power spectrum (dotted line) has been subtracted.}
\label{final_showing_resid.fig}
\end{center}
\end{figure*}

The residual source correction calculated using the T98$^{\star}$ model is applied to the VSA power spectrum as this method does not over-estimate the counts at low flux densities. Integrating the T98$^{\star}$ model up to the subtraction limit of 20~mJy, we get $T_{\rm CMB}^2 C_{\rm src} = 130 \times 10^{-5}$~$\umu$K$^{2}$~sr (or, 207~$\umu$K$^{2}$ at $\ell=1000$, see Table~\ref{resids.table}). This means that the residual source power spectrum is estimated to be $\Delta T_{\rm src}^{2} = 207 \times (\ell/1000)^2$~$\umu$K$^{2}$ and this is binned and subtracted directly from the CMB band-power estimates as shown in  Fig.~\ref{final_showing_resid.fig}.

The T98 30~GHz model must be re-scaled by 0.75 in order to fit the Ka-band WMAP source counts \citep{argueso_etal_03}, and this is consistent at the 1-$\sigma$ level with the re-scaling of $0.68 \pm 0.07$ required by the VSA source counts.

The cosmological parameter estimation incorporating the extended-array power spectrum data of \cite{dickinson_etal_04} was performed by \cite{paper8}. In order to assess the effectiveness of our source subtraction procedure, \cite{paper8} included the parameter $A_{\rm X}$ in the likelihood analysis, where $C_{\ell}=2\pi A_{\rm X} \times 10^{-6}$.  A significant $\Delta T^{2}=A_{\rm X} \ell^{2}$ component of the power spectrum may be considered as an indication of an error in our estimate of the residual source correction assuming that the contribution from the SZ effect is negligible for $\ell \ltsim 1500$.

The combined WMAP and VSA datasets were used for this analysis as well as two external priors, the Hubble Space Telescope Key Project constraint on Hubble's constant of $H_{0}=72 \pm 8$ km sec$^{-1}$ Mpc$^{-1}$ \citep{freedman_etal_01} and constraints on large-scale structure from the 2dF Galaxy Redshift Survey \citep{colless_etal_01}. The analysis gives $A_{\rm X}= -46 \pm 132$ and $A_{\rm X}= -86 \pm 123$ with the HST and 2dF priors respectively. In both cases, $A_{\rm x}$ is consistent with zero and assuming that point sources are the only significant $\Delta T^{2} \propto \, \ell^{2}$ component, the source subtraction has been successful to at least within $\approx 100$~$\umu$K$^{2}$.

\section[]{Implications for other CMB experiments}

\cite{mason_etal_03} have previously found power in excess of the expected primary anisotropy in the multipole range $\ell$=2000--3500 of the TT power spectrum from an analysis of deep field observations with the CBI in 2000. The results of further TT measurements have recently been announced \citep{readhead_etal_04} which, when combined with previous data, act to reduce the significance of the excess power detection. In this section we examine the implications of the 33-GHz source monitoring programme for the original CBI deep-field detection of excess power.

\cite{mason_etal_03} report a band-averaged power 508~$\umu$K$^{2}$ in their highest $\ell$ bin ($2010 < \ell < 3500$) which is 3.1-$\sigma$ in excess of their band-averaged model power of 66~$\umu$K$^{2}$. The detection of excess power could be due to a population of sources with strongly inverted spectra not appearing in NVSS and therefore not `projected-out' of the CBI data or appearing in the residual source correction. All sources detected at 15~GHz and subtracted from the VSA fields had positional counterparts in NVSS except one, which had a strongly inverted spectrum ($\alpha = -2$) between 15 and 33~GHz. There was no evidence, therefore, of a new population of strongly inverted sources at 15~GHz. However it is important to note that, with a lower completeness limit of 20~mJy at 33~GHz, the source count estimate is insensitive to sources with rising spectral indices less than $\alpha \approx -1$. A deeper survey will be required to investigate whether there is a significant population of sources appearing at 33~GHz with spectra more inverted than this.

The excess power could also result from an underestimate of the residual source correction due to sources below the NVSS detection threshold of 3.4~mJy. This correction is estimated to be $C^{\rm res}=0.08 \pm 0.04$~Jy$^{2}$~sr$^{-1}$ by \cite{mason_etal_03} using Monte-Carlo simulations of source populations based on the 31~GHz source counts determined from CBI deep and mosaic maps ($n(S)=92\:S^{-2}$ Jy$^{-1}$ sr$^{-1}$) and the OVRO-NVSS distribution of spectral indices with $\bar{\alpha}_{1.4}^{31}=0.45$. In order to account for the expected flattening of the counts, the re-scaled T98 model was used to determine the residual contribution with the 1.4~GHz limiting flux density of 3.4 m~Jy extrapolated to 33~GHz using the mean NVSS-VSA spectral index $\bar{\alpha}_{1.4}^{33}=0.54$ observed by the VSA. This mean spectral index was derived from the 31 sources in VSA5678 fields with $S_{\rm 33\:GHz} \geq 30$~mJy, in order to ensure that only high signal-to-noise 33-GHz measurements were included. This results in an estimate of $C^{\rm res}=0.03$~Jy$^{2}$ sr$^{-1}$ which is consistent with that found by \cite{mason_etal_03}. It is unlikely, therefore, that Poissonian-distributed point sources constitute the dominant contribution to the CBI excess. However, it is important to note that the observed NVSS-VSA spectral index distribution is subject to various selection effects (discussed in Section~\ref{ssub_src_counts}) which will be addressed by source-subtraction schemes for higher-resolution VSA observations.

\section{Source subtraction for the super-extended VSA}

The VSA has previously observed in configurations probing $\ell$ up to 1500. In principle, observations up to higher $\ell$ are possible, limited by the size of the tilt-table upon which the horn are mounted. The array has now been reconfigured to probe $\ell$ up to 2500. In order to maintain the filling factor of this `super-extended' array, the 322 mm mirrors have been replaced with lightweight 550 mm mirrors. The front-end amplifiers have also been upgraded, resulting in a system temperature decrease from 35 to 28~K.

\begin{figure}
\begin{center}
\includegraphics[scale=0.57, angle=0]{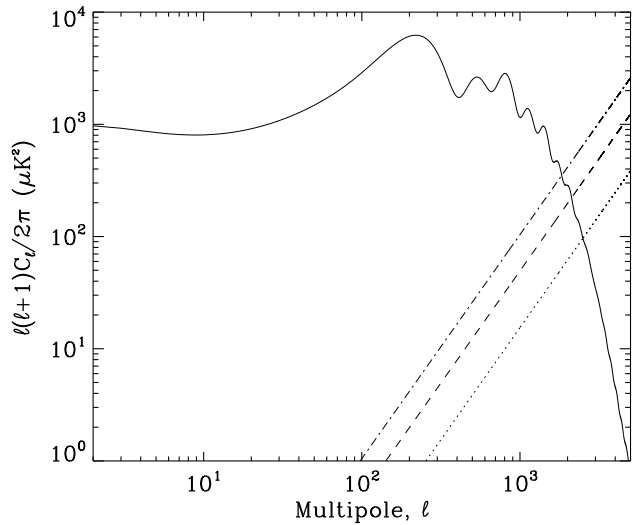}
\caption[]{The residual source power spectrum for a source subtraction limit of 10~mJy (dashed-dot line), 5~mJy (dashed) and 2~mJy (dotted line). Also shown is a model CMB power spectrum. A source subtraction level of $S_{\rm lim} \ltsim 2$~mJy is required for a residual source power spectrum less than the CMB power spectrum at $\ell = 2500$.}
\label{future_ssub.fig}
\end{center}
\end{figure}

Based on the T98$^{\star}$ model, the required source subtraction limit such that the residual source power spectrum is less than a model CMB power spectrum at $\ell=2500$ is $S_{\rm lim} \ltsim 2$~mJy at 33~GHz, as shown in Fig.~\ref{future_ssub.fig}. This deeper source subtraction at 33~GHz also has implications for the 15-GHz survey with the RT. In order that all sources with rising spectra $\alpha=-2$ are found at 5-$\sigma$, the required completion limit at 15~GHz is around 0.4 mJy, requiring a RT sensitivity of $\approx 80$ ${\umu}$Jy. Planned upgrades to the RT as part of the Arcminute Microwave Imager programme \citep{kneissl_etal_01} will result in an increase in flux sensitivity by a factor of $\approx 10$ and hence in survey speed by $\approx 100$, making surveys of the required depth feasible.

An alternative approach involves the use of an instrument currently under development. The goal of the One Centimetre Receiver Array \citep[OCRA;][]{browne_00} programme is to develop a 100-beam receiving system for all-sky surveys at 30~GHz. The first stage (OCRA-p) involves building a 2-beam prototype and this has been installed on the Torun 32 m dish in Poland for testing. The first astronomical task of OCRA-p will be to make an unbiased survey of the discrete sources in the VSA fields down to a few mJy. The brightest sources detected using OCRA-p could then be monitored for variability by the VSA source subtractor during CMB observations by the main array.

\section[]{Conclusions}

The VSA source subtractor has been used to monitor at 33~GHz sources in the VSA fields detected by the Ryle Telescope at 15~GHz. A total of 131 sources with flux densities greater than 20~mJy were subtracted from the data. 

A $\chi^{2}$ test indicated that 8 of the subtracted sources were variable and two of these had well-defined structure functions showing evidence of variability on timescales of days. The source subtractor does not have sufficient instantaneous sensitivity to confirm variability in the fainter sources, however there is evidence of a correlation between the ratio of observed-to-expected r.m.s.\ in the time-ordered flux density data and the spectral index between 1.4 and 33~GHz. The subtracted sources include five new peaked-spectrum candidate sources, with estimated peak frequencies in the range 1--7~GHz.

Based on areas within the VSA fields surveyed by the Ryle Telescope to 10-mJy completeness, the 33-GHz differential source counts were estimated to be $n(S) \approx 21\;S^{-2.34}$ Jy$^{-1}$ sr$^{-1}$. That is, there are estimated to be 0.9 sources per square degree brighter than 20~mJy. The 33-GHz counts presented here are estimated to be complete in the flux density range 20--114~mJy.

Previous VSA analyses have extrapolated the source count to flux densities below the source subtraction limit in order to estimate the residual source contribution. This method takes no account of the expected flattening of the counts at faint flux densities and so is likely to over-estimate the residual contribution. Instead, the 30~GHz differential source-count model of \cite{toff_98} was re-scaled by 0.68 to best fit the VSA data. Integrating the re-scaled T98 model up to 20~mJy provided an estimate of the residual source contribution and this was subtracted from the estimated power spectrum. The residual contribution was estimated to be $\approx 207$~$\umu$K$^{2}$ at $\ell = 1000$. The re-scaling of the T98 model required by the VSA and WMAP source counts indicate that the assumptions about source spectra adopted by \cite{toff_98} may need to be refined.

Using the re-scaled T98 model and the 1.4--33~GHz spectral index statistics, the residual source correction for the CBI deep-field data of \cite{mason_etal_03} was esimated. We find a contribution of 0.03~Jy$^{2}$ sr$^{-1}$ which is consistent with the value of $0.08 \pm 0.04$~Jy$^{2}$~sr$^{-1}$ found by \cite{mason_etal_03}.

\section*{ACKNOWLEDGEMENTS}
The authors made use of the database CATS \citep{Verkhodanov_etal_97} of the Special Astrophysical Observatory. This research has also made use of the NASA/IPAC Extragalactic Database (NED) which is operated by the Jet Propulsion Laboratory, California Institute of Technology, under contract with the National Aeronautics and Space Administration. The authors thank Luigi Toffolatti for making available his numerical 30~GHz source-count predictions. We also thank Tim Pearson and Brian Mason for useful comments on this paper.
\appendix
\section[]{33-GHz Source List}
\label{appendix}

In this section, the 33-GHz source list used for the source count estimates is presented. For each source found to have a flux density greater than 20~mJy at 33 GHz, the measured flux density, $S$, and 1-$\sigma$ thermal error, $\Delta S$, in Jy are presented in Table~A1. In addition to the quoted error, there is a calibration uncertainty of $\approx$ 4 per cent in flux density.

\begin{table*}
\centering
\caption{33-GHz source flux densities.}
\label{slist.table}
\begin{tabular}{lcccccc}
\hline
Name(J2000)&RA(J2000)&Dec(J2000)&$S$ [mJy]&$\Delta S$ [mJy]\\
\hline
0010+3044& 00$^{\rm h}$ 10$^{\rm m}$ 08$\fs$4&30$\degr$ 44$\arcmin$ 42$\arcsec$& 20.0& 3.0\\
0010+2717& 00$^{\rm h}$ 10$^{\rm m}$ 28$\fs$7&27$\degr$ 17$\arcmin$ 54$\arcsec$& 25.8& 6.1\\
0011+2803& 00$^{\rm h}$ 11$^{\rm m}$ 33$\fs$8&28$\degr$ 03$\arcmin$ 46$\arcsec$& 29.4& 6.0\\
0011+2928& 00$^{\rm h}$ 11$^{\rm m}$ 46$\fs$0&29$\degr$ 28$\arcmin$ 28$\arcsec$& 27.3& 3.0\\
0012+2702& 00$^{\rm h}$ 12$^{\rm m}$ 38$\fs$1&27$\degr$ 02$\arcmin$ 40$\arcsec$& 27.5& 6.3\\
0012+3053& 00$^{\rm h}$ 12$^{\rm m}$ 50$\fs$3&30$\degr$ 53$\arcmin$ 19$\arcsec$& 31.5& 3.3\\
0014+2815& 00$^{\rm h}$ 14$^{\rm m}$ 33$\fs$8&28$\degr$ 15$\arcmin$ 08$\arcsec$& 34.8& 3.7\\
0018+2907& 00$^{\rm h}$ 18$^{\rm m}$ 50$\fs$9&29$\degr$ 07$\arcmin$ 40$\arcsec$& 22.9& 2.3\\
0019+2817& 00$^{\rm h}$ 19$^{\rm m}$ 08$\fs$8&28$\degr$ 17$\arcmin$ 56$\arcsec$& 23.9& 2.4\\
0019+2956& 00$^{\rm h}$ 19$^{\rm m}$ 37$\fs$8&29$\degr$ 56$\arcmin$ 02$\arcsec$& 30.9& 2.3\\
0023+2928& 00$^{\rm h}$ 23$^{\rm m}$ 39$\fs$7&29$\degr$ 28$\arcmin$ 22$\arcsec$& 35.5& 2.6\\
0024+2911& 00$^{\rm h}$ 24$^{\rm m}$ 34$\fs$5&29$\degr$ 11$\arcmin$ 29$\arcsec$& 29.8& 2.5\\
0024+2724& 00$^{\rm h}$ 24$^{\rm m}$ 50$\fs$5&27$\degr$ 24$\arcmin$ 27$\arcsec$& 29.7& 4.0\\
0028+3103& 00$^{\rm h}$ 28$^{\rm m}$ 11$\fs$2&31$\degr$ 03$\arcmin$ 35$\arcsec$& 23.0& 5.0\\
0028+2914& 00$^{\rm h}$ 28$^{\rm m}$ 17$\fs$1&29$\degr$ 14$\arcmin$ 29$\arcsec$& 30.2& 4.0\\
0028+2954& 00$^{\rm h}$ 28$^{\rm m}$ 31$\fs$9&29$\degr$ 54$\arcmin$ 52$\arcsec$& 22.4& 3.2\\
0029+2958& 00$^{\rm h}$ 29$^{\rm m}$ 40$\fs$3&29$\degr$ 58$\arcmin$ 51$\arcsec$& 22.0& 3.0\\
0259+2434& 02$^{\rm h}$ 59$^{\rm m}$ 10$\fs$5&24$\degr$ 34$\arcmin$ 12$\arcsec$& 20.0& 5.1\\
0301+2504& 03$^{\rm h}$ 01$^{\rm m}$ 43$\fs$9&25$\degr$ 04$\arcmin$ 43$\arcsec$& 25.3& 3.4\\
0301+2442& 03$^{\rm h}$ 01$^{\rm m}$ 46$\fs$1&24$\degr$ 42$\arcmin$ 58$\arcsec$& 25.0& 5.2\\
0303+2531& 03$^{\rm h}$ 03$^{\rm m}$ 59$\fs$5&25$\degr$ 31$\arcmin$ 35$\arcsec$& 30.0& 2.9\\
0307+2656& 03$^{\rm h}$ 07$^{\rm m}$ 07$\fs$7&26$\degr$ 56$\arcmin$ 07$\arcsec$& 20.6& 2.9\\
0309+2738& 03$^{\rm h}$ 09$^{\rm m}$ 22$\fs$0&27$\degr$ 38$\arcmin$ 56$\arcsec$&113.0& 3.2\\
0310+2810& 03$^{\rm h}$ 10$^{\rm m}$ 18$\fs$8&28$\degr$ 10$\arcmin$ 02$\arcsec$& 28.9& 5.1\\
0311+2530& 03$^{\rm h}$ 11$^{\rm m}$ 40$\fs$9&25$\degr$ 30$\arcmin$ 12$\arcsec$& 55.2& 5.0\\
0313+2718& 03$^{\rm h}$ 13$^{\rm m}$ 44$\fs$9&27$\degr$ 18$\arcmin$ 48$\arcsec$& 30.0& 6.1\\
0314+2636& 03$^{\rm h}$ 14$^{\rm m}$ 58$\fs$3&26$\degr$ 36$\arcmin$ 49$\arcsec$& 41.0& 4.0\\
0708+5409& 07$^{\rm h}$ 08$^{\rm m}$ 30$\fs$5&54$\degr$ 09$\arcmin$ 05$\arcsec$& 20.7& 8.4\\
0712+5430& 07$^{\rm h}$ 12$^{\rm m}$ 44$\fs$6&54$\degr$ 30$\arcmin$ 27$\arcsec$& 23.8& 5.9\\
0714+5343& 07$^{\rm h}$ 14$^{\rm m}$ 47$\fs$4&53$\degr$ 43$\arcmin$ 27$\arcsec$& 29.8& 5.9\\
0715+5609& 07$^{\rm h}$ 15$^{\rm m}$ 16$\fs$8&56$\degr$ 09$\arcmin$ 55$\arcsec$& 22.9& 8.5\\
0716+5323& 07$^{\rm h}$ 16$^{\rm m}$ 40$\fs$4&53$\degr$ 23$\arcmin$ 13$\arcsec$&110.6& 5.9\\
0717+5231& 07$^{\rm h}$ 17$^{\rm m}$ 30$\fs$0&52$\degr$ 31$\arcmin$ 04$\arcsec$& 31.3& 8.9\\
0719+5229& 07$^{\rm h}$ 19$^{\rm m}$ 16$\fs$4&52$\degr$ 29$\arcmin$ 56$\arcsec$& 26.6& 8.8\\
0719+5526& 07$^{\rm h}$ 19$^{\rm m}$ 37$\fs$9&55$\degr$ 26$\arcmin$ 07$\arcsec$& 34.6& 6.2\\
0720+5213& 07$^{\rm h}$ 20$^{\rm m}$ 38$\fs$8&52$\degr$ 13$\arcmin$ 30$\arcsec$& 29.2& 8.8\\
0723+5355& 07$^{\rm h}$ 23$^{\rm m}$ 22$\fs$6&53$\degr$ 55$\arcmin$ 38$\arcsec$& 35.6& 3.3\\
0724+5653& 07$^{\rm h}$ 24$^{\rm m}$ 20$\fs$7&56$\degr$ 53$\arcmin$ 43$\arcsec$& 20.9&13.3\\
0727+5621& 07$^{\rm h}$ 27$^{\rm m}$ 45$\fs$4&56$\degr$ 21$\arcmin$ 14$\arcsec$& 30.2&12.9\\
0728+5325& 07$^{\rm h}$ 28$^{\rm m}$ 03$\fs$3&53$\degr$ 25$\arcmin$ 15$\arcsec$& 52.9& 3.8\\
0728+5431& 07$^{\rm h}$ 28$^{\rm m}$ 23$\fs$8&54$\degr$ 31$\arcmin$ 17$\arcsec$&112.4& 3.8\\
0731+5338& 07$^{\rm h}$ 31$^{\rm m}$ 16$\fs$9&53$\degr$ 38$\arcmin$ 57$\arcsec$& 39.9& 3.6\\
0733+5605& 07$^{\rm h}$ 33$^{\rm m}$ 28$\fs$7&56$\degr$ 05$\arcmin$ 42$\arcsec$& 51.5&12.5\\
0739+5343& 07$^{\rm h}$ 39$^{\rm m}$ 43$\fs$6&53$\degr$ 43$\arcmin$ 07$\arcsec$& 23.2& 4.8\\
0931+3049& 09$^{\rm h}$ 31$^{\rm m}$ 03$\fs$1&30$\degr$ 49$\arcmin$ 12$\arcsec$& 32.6& 9.5\\
0934+3050& 09$^{\rm h}$ 34$^{\rm m}$ 47$\fs$3&30$\degr$ 50$\arcmin$ 52$\arcsec$& 21.9& 3.6\\
0936+3313& 09$^{\rm h}$ 36$^{\rm m}$ 09$\fs$5&33$\degr$ 13$\arcmin$ 09$\arcsec$& 23.8& 3.4\\
0942+3344& 09$^{\rm h}$ 42$^{\rm m}$ 36$\fs$2&33$\degr$ 44$\arcmin$ 37$\arcsec$& 25.0& 4.0\\
0944+3115& 09$^{\rm h}$ 44$^{\rm m}$ 11$\fs$6&31$\degr$ 15$\arcmin$ 24$\arcsec$& 25.3& 3.6\\
0944+3347& 09$^{\rm h}$ 44$^{\rm m}$ 20$\fs$1&33$\degr$ 47$\arcmin$ 56$\arcsec$& 34.8& 3.8\\
0946+3050& 09$^{\rm h}$ 46$^{\rm m}$ 13$\fs$5&30$\degr$ 50$\arcmin$ 22$\arcsec$& 28.5& 3.9\\
0950+3201& 09$^{\rm h}$ 50$^{\rm m}$ 48$\fs$7&32$\degr$ 01$\arcmin$ 43$\arcsec$& 21.1& 5.0\\
1215+5349& 12$^{\rm h}$ 15$^{\rm m}$ 05$\fs$9&53$\degr$ 49$\arcmin$ 55$\arcsec$& 46.0& 5.0\\
1215+5336& 12$^{\rm h}$ 15$^{\rm m}$ 28$\fs$8&53$\degr$ 36$\arcmin$ 08$\arcsec$&114.0&16.0\\
1215+5154& 12$^{\rm h}$ 15$^{\rm m}$ 46$\fs$1&51$\degr$ 54$\arcmin$ 50$\arcsec$& 26.0& 6.0\\
1216+5244& 12$^{\rm h}$ 16$^{\rm m}$ 23$\fs$6&52$\degr$ 44$\arcmin$ 01$\arcsec$& 20.0&10.0\\
1219+5408& 12$^{\rm h}$ 19$^{\rm m}$ 43$\fs$3&54$\degr$ 08$\arcmin$ 33$\arcsec$& 29.0&10.0\\
1221+5429& 12$^{\rm h}$ 21$^{\rm m}$ 18$\fs$2&54$\degr$ 29$\arcmin$ 06$\arcsec$& 20.7& 3.5\\
1223+5409& 12$^{\rm h}$ 23$^{\rm m}$ 13$\fs$1&54$\degr$ 09$\arcmin$ 09$\arcsec$& 34.9& 3.5\\
1229+5522& 12$^{\rm h}$ 29$^{\rm m}$ 09$\fs$3&55$\degr$ 22$\arcmin$ 30$\arcsec$&109.2& 8.4\\
1229+5147& 12$^{\rm h}$ 29$^{\rm m}$ 22$\fs$0&51$\degr$ 47$\arcmin$ 07$\arcsec$& 41.3& 4.9\\
1234+5054& 12$^{\rm h}$ 34$^{\rm m}$ 16$\fs$0&50$\degr$ 54$\arcmin$ 23$\arcsec$& 33.2& 5.4\\
1235+5228& 12$^{\rm h}$ 35$^{\rm m}$ 30$\fs$6&52$\degr$ 28$\arcmin$ 28$\arcsec$& 27.9& 2.6\\
1235+5340& 12$^{\rm h}$ 35$^{\rm m}$ 48$\fs$0&53$\degr$ 40$\arcmin$ 05$\arcsec$& 62.1& 3.5\\
\hline
\end{tabular}
\end{table*}

\begin{table*}
\centering
\contcaption{33-GHz source flux densities.}
\label{slist1.table}
\begin{tabular}{lcccccc}
\hline
Name(J2000)&RA(J2000)&Dec(J2000)&$S$ [mJy]&$\Delta S$ [mJy]\\
\hline
1237+5057& 12$^{\rm h}$ 37$^{\rm m}$ 23$\fs$6&50$\degr$ 57$\arcmin$ 21$\arcsec$& 79.0& 6.0\\
1238+5325& 12$^{\rm h}$ 38$^{\rm m}$ 07$\fs$6&53$\degr$ 25$\arcmin$ 54$\arcsec$& 27.2& 3.6\\
1240+5334& 12$^{\rm h}$ 40$^{\rm m}$ 05$\fs$5&53$\degr$ 34$\arcmin$ 33$\arcsec$& 27.1& 6.2\\
1240+5441& 12$^{\rm h}$ 40$^{\rm m}$ 14$\fs$2&54$\degr$ 41$\arcmin$ 48$\arcsec$& 27.8& 8.8\\
1240+5334& 12$^{\rm h}$ 40$^{\rm m}$ 42$\fs$4&53$\degr$ 34$\arcmin$ 23$\arcsec$& 24.8& 7.0\\
1241+5141& 12$^{\rm h}$ 41$^{\rm m}$ 16$\fs$5&51$\degr$ 41$\arcmin$ 29$\arcsec$& 22.0& 5.0\\
1523+4156& 15$^{\rm h}$ 23$^{\rm m}$ 09$\fs$4&41$\degr$ 56$\arcmin$ 25$\arcsec$& 27.9& 4.3\\
1523+4420& 15$^{\rm h}$ 23$^{\rm m}$ 34$\fs$4&44$\degr$ 20$\arcmin$ 31$\arcsec$& 23.9&10.9\\
1526+4201& 15$^{\rm h}$ 26$^{\rm m}$ 45$\fs$3&42$\degr$ 01$\arcmin$ 41$\arcsec$& 34.2& 3.9\\
1528+4219& 15$^{\rm h}$ 28$^{\rm m}$ 00$\fs$1&42$\degr$ 19$\arcmin$ 14$\arcsec$& 27.4& 2.5\\
1533+4107& 15$^{\rm h}$ 33$^{\rm m}$ 27$\fs$9&41$\degr$ 07$\arcmin$ 23$\arcsec$& 28.1& 3.3\\
1539+4217& 15$^{\rm h}$ 39$^{\rm m}$ 25$\fs$7&42$\degr$ 17$\arcmin$ 27$\arcsec$& 34.0& 2.3\\
1539+4123& 15$^{\rm h}$ 39$^{\rm m}$ 36$\fs$8&41$\degr$ 23$\arcmin$ 33$\arcsec$& 33.7& 3.3\\
1541+4114& 15$^{\rm h}$ 41$^{\rm m}$ 01$\fs$2&41$\degr$ 14$\arcmin$ 28$\arcsec$& 32.5& 3.2\\
1545+4130& 15$^{\rm h}$ 45$^{\rm m}$ 21$\fs$4&41$\degr$ 30$\arcmin$ 27$\arcsec$& 30.2& 4.0\\
1546+4449& 15$^{\rm h}$ 46$^{\rm m}$ 04$\fs$6&44$\degr$ 49$\arcmin$ 13$\arcsec$& 24.0& 2.0\\
1546+4257& 15$^{\rm h}$ 46$^{\rm m}$ 22$\fs$5&42$\degr$ 57$\arcmin$ 56$\arcsec$& 23.6& 4.1\\
1723+4206& 17$^{\rm h}$ 23$^{\rm m}$ 25$\fs$1&42$\degr$ 06$\arcmin$ 10$\arcsec$& 24.4& 3.4\\
1726+3957& 17$^{\rm h}$ 26$^{\rm m}$ 32$\fs$6&39$\degr$ 57$\arcmin$ 02$\arcsec$&109.5& 6.3\\
1727+4221& 17$^{\rm h}$ 27$^{\rm m}$ 49$\fs$3&42$\degr$ 21$\arcmin$ 41$\arcsec$& 44.3& 2.6\\
1728+3935& 17$^{\rm h}$ 28$^{\rm m}$ 32$\fs$1&39$\degr$ 35$\arcmin$ 16$\arcsec$& 24.1& 6.3\\
1730+4102& 17$^{\rm h}$ 30$^{\rm m}$ 41$\fs$4&41$\degr$ 02$\arcmin$ 57$\arcsec$& 39.3& 2.7\\
1731+4101& 17$^{\rm h}$ 31$^{\rm m}$ 23$\fs$6&41$\degr$ 01$\arcmin$ 38$\arcsec$& 25.8& 2.2\\
1733+4034& 17$^{\rm h}$ 33$^{\rm m}$ 07$\fs$5&40$\degr$ 34$\arcmin$ 28$\arcsec$& 20.8& 6.2\\
1734+3857& 17$^{\rm h}$ 34$^{\rm m}$ 20$\fs$5&38$\degr$ 57$\arcmin$ 52$\arcsec$&939.3& 1.9\\
1737+3958& 17$^{\rm h}$ 37$^{\rm m}$ 41$\fs$1&39$\degr$ 58$\arcmin$ 18$\arcsec$& 36.1& 6.4\\
1737+4154& 17$^{\rm h}$ 37$^{\rm m}$ 59$\fs$4&41$\degr$ 54$\arcmin$ 54$\arcsec$& 28.1& 2.4\\
1738+4008& 17$^{\rm h}$ 38$^{\rm m}$ 19$\fs$1&40$\degr$ 08$\arcmin$ 20$\arcsec$&105.7& 6.4\\
1738+4026& 17$^{\rm h}$ 38$^{\rm m}$ 49$\fs$7&40$\degr$ 26$\arcmin$ 21$\arcsec$& 20.6& 6.4\\
1740+4234& 17$^{\rm h}$ 40$^{\rm m}$ 51$\fs$9&42$\degr$ 34$\arcmin$ 47$\arcsec$& 57.0& 5.0\\
1744+4014& 17$^{\rm h}$ 44$^{\rm m}$ 25$\fs$1&40$\degr$ 14$\arcmin$ 49$\arcsec$& 97.3& 6.1\\
1745+4059& 17$^{\rm h}$ 45$^{\rm m}$ 28$\fs$5&40$\degr$ 59$\arcmin$ 52$\arcsec$&267.7& 6.2\\

\hline
\end{tabular}
\end{table*}

\bibliographystyle{mn2e}
\bibliography{litrefs}

\end{document}